\documentclass[10pt]{iopart}
\usepackage{iopams}
\usepackage{graphicx}
\usepackage[usenames,dvipsnames]{color}
\usepackage{cite}

\newtheorem{theorem}{Theorem}
\input{epsf}

\begin{document}

\title{Exploring the Gillis model: a discrete approach to diffusion in logarithmic potentials}
\author{Manuele Onofri$^{1,2}$, Gaia Pozzoli$^{1,2}$, Mattia Radice$^{1,2}$, Roberto Artuso$^{1,2}$}
\address{$^1$ Dipartimento di Scienza e Alta Tecnologia and Center for Nonlinear and Complex Systems, Universit\`a degli Studi dell'Insubria, Via Valleggio 11, 22100 Como Italy}
\address{$^2$ I.N.F.N. Sezione di Milano, Via Celoria 16, 20133 Milano, Italy}
\eads{\mailto{m.onofri1@uninsubria.it}, \mailto{gpozzoli@uninsubria.it}, \mailto{m.radice1@uninsubria.it, \mailto{roberto.artuso@uninsubria.it}}}
\vspace{10pt}

\begin{abstract}
Gillis model, introduced more than 60 years ago, is a non-homogeneous random walk with a position dependent drift. Though parsimoniously cited both in the physical and mathematical literature, it provides one of the very few examples of a stochastic system allowing for a number of exact result, although lacking translational invariance. We present old and novel results for such model, which moreover we show represents a discrete version of a diffusive particle in the presence of a logarithmic potential.
\end{abstract}

\vspace{2pc}
\noindent{\it Keywords }: Gillis random walk, Logarithmic potential, Anomalous transport, Ergodicity, Maxima \\

\section{Introduction}\label{Sec:1}
The \textit{Gillis model}  is a centrally biased random walk on an integer lattice introduced in 1956 \cite{Gillis} to study the recurrence properties of a random process with transition probabilities lacking translational invariance. In one dimension the model consists in a walker that starts its motion in $j_0=0$ and moves on the integer lattice making jumps only between first neighbour sites. The probabilities  $\mathcal{R}(j)$ and $\mathcal{L}(j)$ of making a jump to the right or to the left depend on the current site $j$ in the following way:
\begin{equation}
\mathcal{R}(j)=\frac{1}{2}\left(1- \frac{\epsilon}{j}\right)\quad \textrm{and} \quad \mathcal{L}(j)=\frac{1}{2}\left(1+ \frac{\epsilon}{j}\right),
\label{Gillis:1d_1}
\end{equation}
for $j\neq 0$, and
\begin{equation}
\mathcal{R}(0)=\mathcal{L}(0)=\frac{1}{2}
\label{Gillis:1d_2}
\end{equation}
when the particle is at the origin; in the previous equations the parameter $\epsilon$ takes its value in the range $(-1,1)$. Consequently for positive values of $\epsilon$ the walker is biased towards the origin, while for negative values it tends to escape from it, while for $\epsilon=0$ one is back to the simple symmetric random walk in one dimension. Thus, except for the trivial case $\epsilon=0$, this model represents a non-homogeneous random walk symmetric with respect to the starting site $j_0=0$.

The mathematical relevance of the model lies in the fact that it is one of few examples of non-homogeneous random walk that can be solved analytically. For example in the original paper by Gillis \cite{Gillis} the author is able to find the exact form of the generating function of the probability of being at the origin after $n$ steps in the one-dimensional case. Other results can be deduced such as the probability of eventual return to the starting site as well as the mean number of steps occurring between two consecutive visits at the origin \cite{Hughes-1986}. However, the result that has attracted more interest regards the recurrence of the starting site (a point is defined recurrent if the walk will, with probability $1$, pass an infinite number of times through it) depending both on the parameter $\epsilon$ and the dimensions $d$ of the system. In this respect the Gillis random walk was the starting point for many papers, see \cite{Lamperti-1960, Lamperti-1963, Hry-Men-Wade, Nash-Williams, Mens-Pop-Wade}, concerning recurrence and transience, maxima and passage-time moments of stochastic processes.

In the physical literature Gillis model attracted little interest apart from \cite{Chan-Hughes}, where the problem of ions diffusion in a semi-infinite domain in the presence of a charged barrier is mapped to the one-dimensional Gillis random walk on the positive set of integers with $\epsilon=1$.
Here we will generalize the result obtained in \cite{Chan-Hughes} and demonstrate that in the continuum limit the Gillis random walk (GRW)  corresponds to the diffusion of a particle in the presence of a logarithmic potential tuned by the parameter $\epsilon$. This problem has been studied in depth, see for example \cite{Dechant,Hirschberg}, as it has been recognized as a natural model for a large number of physical systems, from vortex dynamics \cite{Bray}, to interaction between probe particles in a driven fluid \cite{Lev-Muk-Sch}, to time evolution of momenta of cold atoms trapped in optical lattices \cite{Castin, Mark-Ell-Zoll, Lutz, Doug-Ger-Renz}, to relaxation to equilibrium of long-range interacting systems \cite{Bouch-Daux,Campa-Daux}. Moreover, diffusion in an effective logarithmic potential also appears outside the physical context, such as the study of charged particles in the vicinity of a charged polymer \cite{Manning}, dynamics of DNA denaturation \cite{Bar-Kaf-Muk} and sleep-wake transitions during sleep \cite{Lo2002}.

In this paper we will consider the one-dimensional Gillis model. Firstly (Sec.\ref{Sec:2}), in order to provide an overview of the problem to the reader, we will present a number of results already known in the literature regarding both local and non-local properties of the stochastic process, such as the first return probability \cite{Hughes-1986}, the distribution of the occupation time of the positive axis and the distribution of the number of returns at the origin \cite{ROAP}. Furthermore, we will present Lamperti criteria \cite{Lamperti-1960, Lamperti-1963}, through which it is possible to determine asymptotic properties of a certain class of stochastic processes. Moreover, we will also show that the process is ergodic for $\epsilon\in\left( \frac{1}{2},1\right)$ by computing the stationary distribution. Afterwards (Sec.\ref{Sec:3}), making use of an appropriate continuum limit, we will demonstrate that the GRW is equivalent to the diffusion of particles in a logarithmic potential tuned by the parameter $\epsilon$. This result enables us to obtain the entire moments' spectrum and shows that the process presents two different phases: a non-ergodic phase, where transport is normal, and an ergodic phase characterized by a strongly anomalous subdiffusion. In Section \ref{Sec:4} we provide the asymptotic behaviour of the mean value of the maximum and show that also in this case the Gillis model presents a phase transition. Finally, we introduce a new model, which consists in a generalized version of the Gillis one, and prove that it is equivalent to a particle diffusing in a power-law potential (Sec.\ref{Sec:5}).

\section{The original model, known results, the stationary distribution and time statistics}\label{Sec:2}
In this section we consider the one-dimensional model, see eqs.\eref{Gillis:1d_1} and \eref{Gillis:1d_2}, presented by Gillis in \cite{Gillis} and mainly provide results known so far in literature with the aim of giving the reader a complete overview of the problem. 

For this process it is possible to obtain the exact expression for the generating function $P(z)$ of the probability $P_n$ of finding the particle at the origin, which is also the starting site, after $n$ steps, see \cite{Gillis}:
\begin{equation}
P(z)=\sum_{n=0}^\infty P_n z^n=\frac{_2 F_1\left(\frac{1}{2}\epsilon+1,\frac{1}{2}\epsilon+\frac{1}{2};1;z^2\right)}{_2 F_1\left(\frac{1}{2}\epsilon,\frac{1}{2}\epsilon+\frac{1}{2};1;z^2\right)},
\label{Gillis:P(z)}
\end{equation}
where $_2 F_1(a,b;c;z)$ is the gaussian hypergeometric function \cite{Abr-Steg}. Actually it is possible to generalize the Gillis solution by considering an arbitrary starting site $j_0$, this result is presented in \ref{App_A}. The knowledge of this generating function allows to derive a number of important properties of the random walk.

\subsection{Probability of being at the origin}\label{Sub_Sec:Pn}
The first quantity that can be obtained quite straightforwardly from $P(z)$ is the asymptotic behaviour of the probability $P_n$. In fact, by using the properties of the hypergeometric function $_2 F_1(a,b;c;z)$ \cite{Abr-Steg}, it is possible to demonstrate that $P(z)$ has the following form (see \ref{App_B}):
\begin{equation}
P(z)=\frac{1}{(1-z)^\nu}H\left(\frac{1}{1-z}\right),
\label{P(z):Form}
\end{equation}
where $H(x)$ is a slowly varying function and $\nu$ assumes the values:
\begin{equation}
\nu= \cases{
0 & for $ -1<\epsilon\leq-\frac{1}{2}$\\
\frac{1}{2}+\epsilon & for $ -\frac{1}{2}<\epsilon\leq\frac{1}{2}$\\
1 & for $\frac{1}{2}<\epsilon<1$.
}
\label{P(z):rho}
\end{equation}
Therefore, $P(z)$ fulfils the Tauberian theorem for power series \cite{Feller} and one obtains:
\begin{equation}
P_{2n}\sim \cases{
n^{-\frac{1}{2}+\epsilon} & for $-1<\epsilon<\frac{1}{2}$\\
\frac{4}{\log n} & for $\epsilon=\frac{1}{2}$,
}
\label{P_n:asympt}
\end{equation}
while for $\frac{1}{2}<\epsilon<1$ the probability $P_{2n}$ converges asymptotically to a constant:
\begin{equation}
P_{2n} \to 2-\frac{1}{\epsilon}.
\label{P_n:asympt:2}
\end{equation}
Obviously, in all cases one has $P_{2n+1}=0$ due to the fact that the walker must perform an even number of steps to reach the starting point.

We underline that if we considered the generating function $P(z|j_0)$ (see \ref{App_A}) of the probability $P_n(0|j_0)$ of being in $n$ steps at the origin having started from the site $j_0$ instead of $P(z)$, we would obtain that the asymptotics of the probability $P_n(0|j_0)$ are the same of $P_n$. However, we note that a walker can reach the origin only in a number of steps whose parity is that of $j_0$, thus the $2n$ index must be replaced by $2n + |j_0|$. For instance, one has that for $\epsilon>\frac{1}{2}$ and any starting site $j_0$
\begin{equation}\label{Pn_k0}
P_{2n+|j_0|}(0|j_0)\to 2-\frac{1}{\epsilon}.
\end{equation}

\subsection{Probability of first return to the origin and recurrence}
Now we consider the probability $F_n$ that the particle returns to the starting point for the first time at the $n$-th step. In order to find the asymptotic behaviour of $F_n$ one can use again the generating function
\begin{equation}
F(z)=\sum_{n=0}^\infty F_n z^n,
\end{equation}
which is connected to $P(z)$ by \cite{Redner}
\begin{equation}
F(z)=1-\frac{1}{P(z)}.
\label{F(z):P(z)}
\end{equation}
 Thus, making use of eq.\eref{P(z):Form},  the generating function of the first return probability is of the following form:
\begin{equation}
F(z)=1-(1-z)^\nu L\left(\frac{1}{1-z} \right),
\label{F(z):form}
\end{equation}
where $L(z)=1/H(z)$. We report the details of the calculation of $F_n$ in \ref{App_C} and here we give only the results regarding its asymptotic behaviour:
\begin{equation}
F_{2n}\sim \cases{
n^{-(1/2-\epsilon)} ~~~& for $-1<\epsilon<-\frac{1}{2}$\\
\frac{1}{n\log^2(n)} & for $\epsilon=-\frac{1}{2}$\\
n^{-(3/2+\epsilon)}~~~& for $-\frac{1}{2}<\epsilon<1$.\\
}
\label{1D:Fn:exp}
\end{equation}
In \ref{App_C} we also calculate the mean recurrence time $\tau_n$, namely the mean time occurring between two consecutive visits at the starting site in a $n$ steps walk, and it is given by, see also \cite{Hughes-1986}: 
\begin{equation}
\tau_n\sim \cases{
n^{3/2+\epsilon} ~~~&for $-1<\epsilon<-\frac{1}{2}$\\
\frac{n}{\log^2(n)} &for $\epsilon=-\frac{1}{2}$\\
n^{1/2+\epsilon}~~~&for $-\frac{1}{2}<\epsilon<\frac{1}{2}$\\
\log(n), &for $\epsilon=\frac{1}{2}$\\
\frac{2\epsilon}{2\epsilon-1}&for $\frac{1}{2}<\epsilon<1$.
}
\label{tau}
\end{equation}
We underline that the Gillis random walk for $\frac{1}{2}<\epsilon<1$ is characterized by a finite mean recurrence time, while in the other cases it increases with the number of steps.

An important quantity used to describe the stochastic processes is the return probability $R$, i.e. the probability that the walker returns to the starting point, which is given by
\begin{equation}
\mathrm{R}=\sum_{n=1}^\infty F_n=\left.F(z)\right|_{z=1}.
\end{equation}
From the relation \eref{F(z):P(z)} it arises that a necessary and sufficient condition for recurrence is the divergence of $P(z)$ for $z\to 1^-$, see \cite{Hughes-1986}. In our case, taking into account the expression for $P(z)$ written previously and considering the properties of the hypergeometric function, one obtains
\begin{equation}\label{Recurrence/Transience}
\mathrm{R}=\cases{
 \vert \epsilon\vert^{-1}-1~~~&for $-1<\epsilon<-\frac{1}{2}$\\
1&for $-\frac{1}{2}\leq\epsilon<1$. 
}
\end{equation}
So the model exhibits transience for $-1<\epsilon<-\frac{1}{2}$, while for $-\frac{1}{2}\leq\epsilon<1$ the process is recurrent. 

\subsection{Lamperti criteria for stochastic processes}\label{Sec:Lamperti:Criteria}
We observed that the recurrence/transience of a stochastic process can be obtained straightforwardly by evaluating the generating function $F(z)$ of the first return probability at $z=1$ or, equivalently, taking the limit $z\to 1$ of the generating function $P(z)$ of the probability of being in the origin. However, the classical approaches to compute these quantities, such as combinatorial ones, are of limited use even if the model is slightly modified (to this purpose in Section \ref{Sec:5} we will present a generalization of the Gillis model that can be easily studied through the theorems presented below). 

In two pioneering papers \cite{Lamperti-1960, Lamperti-1963} Lamperti, in order to study the asymptotic behaviour of stochastic processes, suggested a method based on the so called \textit{Lyapunov functions} (see an example in \ref{App_D}). The results obtained by Lamperti consist, essentially, in finding some criteria that the moments of the increment must satisfy to determine the asymptotics of the process. In the following we will briefly present the criteria that can be applied to the Gillis model; since we consider Lamperti criteria important to enhance the understanding of stochastic processes, the following part will be as consistent as possible with the original papers.

First of all let us describe the quantities of interest. Let $\{ X_n \}$, with $n\in \mathbb{N}$, be a Markov process in $\mathbb{R}_+$ with stationary transition probabilities and define with $\mu_k(x)$ the $k$-th moment of the increment $\Delta_n=X_{n+1}-X_n$ given $X_n=x$:
\begin{equation}
\mu_k(x)=\mathbb{E}\left[ (X_{n+1}-X_n)^k|X_n=x \right].
\end{equation}
For the time being, let us suppose that $\mu_k(x)$ is well defined for all $k$, for example imposing uniformly boundedness of the increments $\Delta_n$, which is obviously verified in the case of a nearest neighbour random walk on the integer lattice. Anyway, we will see that the above request is crucial only for the first two moments. 

The first theorem, see \cite{Lamperti-1960}, concerns the transience/recurrence of the stochastic process $\{X_n\}$:
\begin{theorem}\label{Th:Lamp:Rec/Trans}
Let $\mu_2(x)$ be bounded away from 0. Suppose that for sufficiently large enough $x$,
\begin{equation}\label{Criteria:Rec}
\mu_1(x)\leq \frac{\theta \mu_2(x) }{2 x}
\end{equation}
for some $\theta<1$. Then $\{X_n\}$ is recurrent. Conversely, if for sufficiently large $x$ and some $\theta>1$
\begin{equation}\label{Criteria:Trans}
\mu_1(x)\geq \frac{\theta \mu_2(x) }{2 x}
\end{equation}
then $\{X_n\}$ is transient ($X_n\to \infty$ a.s.).
\end{theorem}
Therefore, given the first two moments $\mu_1(x)$ and $\mu_2(x)$ for the Gillis model
\begin{eqnarray}\label{mu:1} 
\mu_1(x)=(+1)\cdot\mathcal{R}(x)+(-1)\cdot\mathcal{L}(x)=-\frac{\epsilon}{x} \\ 
\label{mu:2}
\mu_2(x)=(+1)^2\cdot\mathcal{R}(x)+(-1)^2\cdot\mathcal{L}(x)=1,
\end{eqnarray} 
one has that \eref{Criteria:Rec} and \eref{Criteria:Trans}  are satisfied respectively for $\epsilon>-\frac{1}{2}$ and  $\epsilon<-\frac{1}{2}$, which is in complete agreement with \eref{Recurrence/Transience}. The limiting case $\epsilon=-\frac{1}{2}$ is treated in a more general theorem (Theorem 3.2 in \cite{Lamperti-1960}) and results to be recurrent.

Further theorems regard the existence or not of the passage-time moments, i.e. the moments of the first return probability $F_n$. Now, let $\lbrace X_n\rbrace$ be a discrete-time stochastic process on a Borel subset of the non-negative reals and we assume that it is Markovian with stationary probabilities (this requirement can be relaxed, see \cite{Lamperti-1963} for details).  For the existence of the passage-time moments one has:
\begin{theorem}\label{Th:Lamp:Finite}
Suppose there exists $\varepsilon>0$ and $A<\infty$ such that, for $x\geq A$, $\mu_2(x)$ exists and
\begin{equation}
2x\mu_1(x)+\mu_2(x)\leq -\epsilon.
\end{equation}
Let $T\geq 0$ be the time at which the process first enters the interval $[0,A]$. Then
\begin{equation}
\mathbb{E}(T)\leq\frac{\mathbb{E}(X_0^2)}{\varepsilon}.
\end{equation}
\end{theorem}
While for the non-existence:
\begin{theorem}
Suppose that the conditional moments $\mu_1(x)$ and $\mu_2(x)$ satisfy
\begin{equation}
2x\mu_1(x)+\mu_2(x)\geq \varepsilon>0
\end{equation}
for all $x\geq A$, and in addition that 
\begin{equation}
\mu_1(x)=O(x^{-1}), \quad \mu_2(x)=O(1), \quad \mu_4(x)=o(x^2).
\end{equation}
Then the time $T_{x_0}$ of first passage from $x_0>A$ to $[0,A]$ has infinite expectation.
\end{theorem}    
From the theorems above one has that in the Gillis process the mean recurrence time is finite only for $\epsilon>\frac{1}{2}$, which is the same result obtained before, see eq.\eref{tau}.

We observe that non-homogeneous random walks, such as the Gillis model and the generalized one treated at the end of this paper, provide an appropriate environment in order to investigate the critical behaviour in the proximity of a phase transition. In fact, these kind of processes can display anomalous recurrence behaviour with respect to the spatially homogeneous ones: while keeping fixed the dimensionality of the model, one can observe either transience or recurrence property by simply changing the parameter value. This fact contrasts with the generalization of P\'olya's theorem \cite{Chung-Fuchs} on the recurrence of spatially homogeneous random walks in $d$-dimensions, which is recurrent for $d\leq 2$ and transient for $d>2$. A very nice example is provided by elliptic random walks \cite{Mens-Pop-Wade, RWell}.

To summarize, if we deal with a one-dimensional random walk characterized by $\mu_1(x) \sim \frac{a}{x}$ and $\mu_2(x) \sim b$, one can define the key parameter $m = \frac{2a}{b^2}$ ($m=-2\epsilon$ in the Gillis random walk) and it follows that $\{X_n \}$ is, see also \cite{Hry-Men-Wade}:
\begin{itemize}
\item \textit{transient} if $m<-1$;
\item \textit{null-recurrent} (the process is recurrent and the mean first passage time is infinite) if $-1\leq m \leq 1 $;
\item \textit{positive-recurrent} (the process is recurrent and the mean first passage time is finite) if $ m>1$.
\end{itemize}

\subsection{Stationary solution}\label{Sec:Stationary:Solution}
In this section we will prove the existence of a stationary distribution for the Gillis random walk with $\epsilon$ in the interval $ \left(\frac{1}{2} ,1 \right)$.
In general, the stationary solution of a random walk on a state space $S$ characterized by transition probabilities $t(j|i)$ of moving from site $i$ to site $j$ is a set of non-negative numbers $\lbrace \pi_j : j\in S \rbrace$ such that:
\begin{eqnarray}\label{St:1} 
\sum_{j\in S} \pi_j=1, \\ 
\label{St:2}
\sum_{i} t(j|i) \pi_i=\pi_j,
\end{eqnarray} 
namely $\pi_j$ are the components of an eigenvector of the transition matrix $t_{ji} = t(j|i)$ with eigenvalue $1$. In our case transitions probabilities are nonzero only for jumps between first neighbour sites and are given by $\mathcal{R}(i)$ and $\mathcal{L}(i)$ (see eqs.\eref{Gillis:1d_1}-\eref{Gillis:1d_2}). Due to the symmetry of GRW, the stationary distribution must satisfy the  symmetry condition $\pi_j=\pi_{-j}$, so that we have to solve the infinite set of linear equations:
\begin{equation}
\cases{
\pi_0=2 \mathcal{L}(1) \pi_1\\
\pi_1=\mathcal{R}(0) \pi_0+\mathcal{L}(2) \pi_2\\
\vdots\\
\pi_j=\mathcal{R}(j-1) \pi_{j-1}+\mathcal{L}(j+1) \pi_{j+1}\\
\vdots\\
}
\label{delta:lamp}
\end{equation}
By iteration one obtains:
\begin{equation}
\pi_j=\frac{j (1-\epsilon)_{j-1}}{(1+\epsilon)_{j}} \pi_0, \quad j\geq 1,
\end{equation}
where $(x)_n$ is the Pochhammer's symbol \cite{Abr-Steg} and $\pi_0$ can be determined by the normalization condition \eref{St:1}. If the distribution $\pi_j$ can not be normalized, then it does not represent a proper distribution of the process and we say that the walk does not admit a stationary distribution. 

With the aim of determining $\pi_0$, we firstly evaluate the behaviour of $\pi_j$ at large distances from the origin; by using the definition of $(x)_n$ in terms of the Gamma function, i.e. $(x)_n=\Gamma(x+n)/\Gamma(x)$, we may deduce that
\begin{equation}\label{Gillis:ST}
\pi_j\sim \pi_0 \frac{\Gamma(1+\epsilon)}{\Gamma(1-\epsilon)} j^{-2\epsilon}.
\end{equation}
Therefore, we observe that the Gillis random walk admits a stationary solution only for $\epsilon>\frac{1}{2}$ and we can say that in this range the process is ergodic. 

Now, instead of using the normalizing condition \eref{St:1}, we obtain $\pi_0$ by demonstrating that it must be equal to the inverse of the mean recurrence time at the origin, which we have seen in \eref{tau} to be finite in the ergodic range. Firstly, let us define with $j_n$ the position reached in $n$ steps by the walker started from $j_0$ and with $V_n$ the number of visits at the origin, which is given by
\begin{equation}
V_n=\sum_{i=1}^n \delta_{0,j_i}.
\end{equation}
From the Birkhoff's ergodic theorem one has that the time average of $V_n$ converges asymptotically to the ensemble average of $\delta_{0,j}$, which is given by the stationary probability $\pi_0$, thus:
\begin{equation}\label{Ret:Birkhoff}
\lim_{n\to \infty} \frac{V_n}{n}=\pi_0.
\end{equation}
Moreover, given a walk of $n$ steps and knowing that the walker visited the origin $V_n$ times, we can evaluate the mean recurrence time $\tau$ between two visits by simply taking the ratio between $n$ and $V_n$, so it holds:
\begin{equation} \label{Ret:Time}
\lim_{n\to \infty} \frac{V_n}{n}=\frac{1}{\tau}.
\end{equation}
Finally, comparing \eref{Ret:Birkhoff} and \eref{Ret:Time} one has that $\pi_0$ and $1/\tau$ must be equal, namely
\begin{equation}\label{St:p(0)}
\pi_0=\frac{2\epsilon-1}{2\epsilon}.
\end{equation}

\begin{figure*}[h!]
	\centering
	\begin{tabular}{c @{\quad} c }
		\includegraphics[width=.45\linewidth]{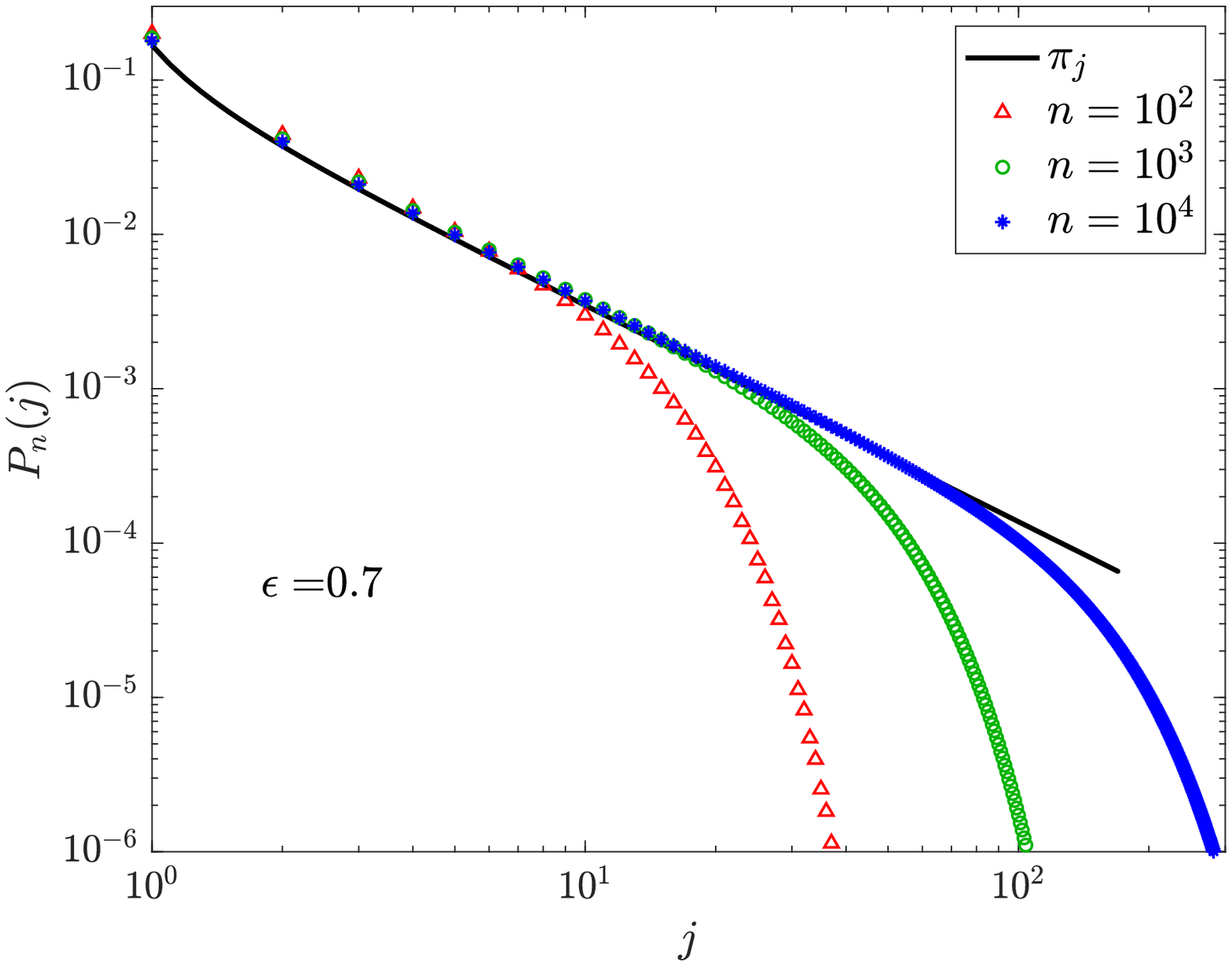} &
		\includegraphics[width=.45\linewidth]{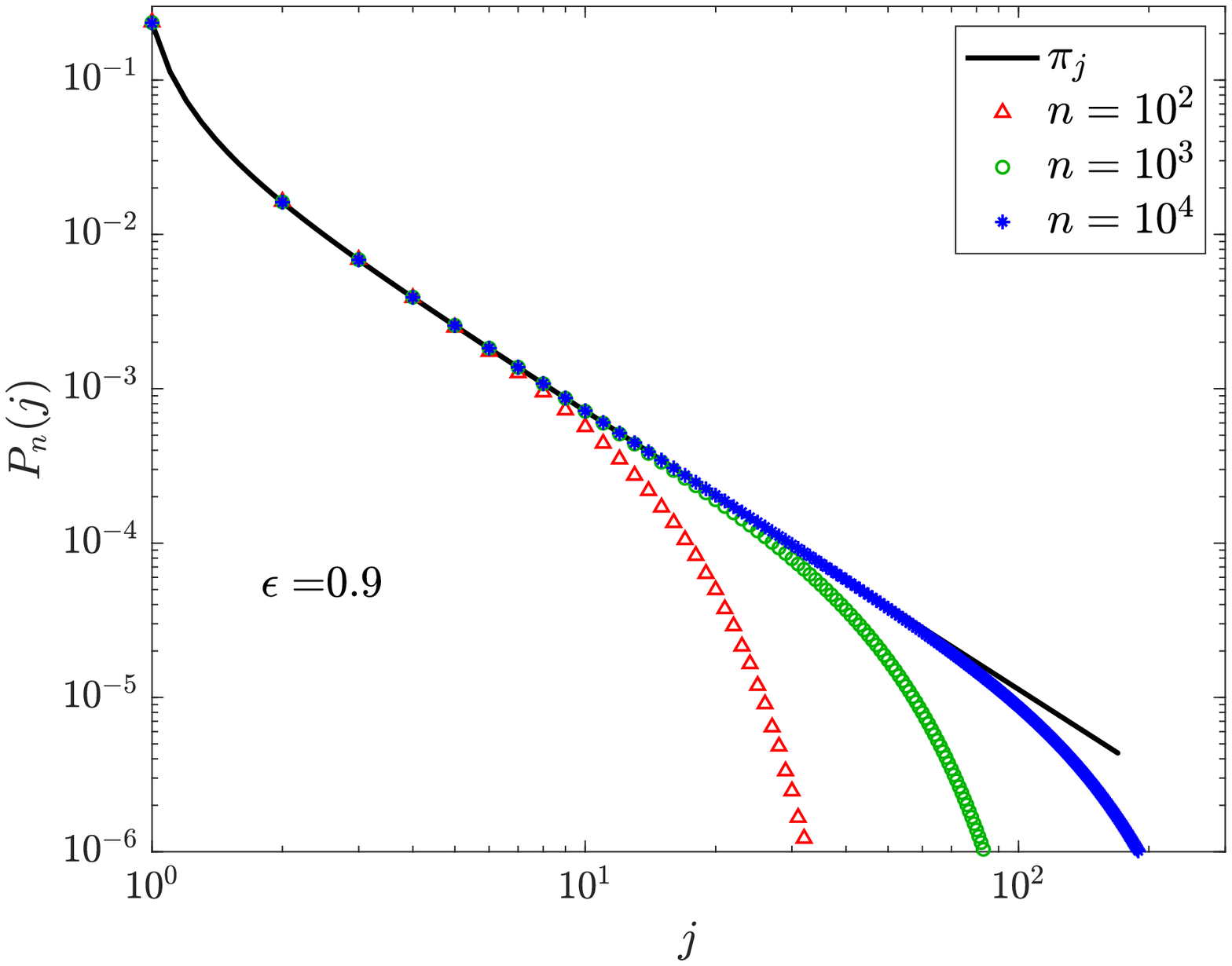} \\
		\small (a) & \small (b)
	\end{tabular}	
	\caption{Comparison between the stationary distribution $\pi_j$ and the probability $P_n(j)$ of finding the particle at site $j$ after $n$ steps for $\epsilon=0.7$ and $0.9$. The evolution of the probability $P_n(j)$ is obtained by considering the master equation \eref{Pdf:master:eq}, presented in section \ref{Sec:3}, with initial condition $P_0(j)=\frac{1}{2}\delta_{0,j}+\frac{1}{2}\delta_{1,j}$. The markers present the distribution $P_n(j)$ for different number of steps (triangles for $n=10^2$, circles for $n=10^3$ and squares for $n=10^4$), while the line shows the stationary distribution $\pi_j$ given in \eref{Gillis:ST}.}
	\label{fig:St}
\end{figure*}

In figure \ref{fig:St} we show the comparison between the stationary distribution $\pi_j$ and the probability $P_n(j)$ of finding the particle at site $j$ after $n$ steps. We observe that $P_n(j)$ tends for large number of steps to the stationary one. However, in fig.\ref{fig:St}(a), we observe that the convergence is slower as the value of $\epsilon$ is nearer to the limiting case $\epsilon=0.5$.
 
\subsection{Occupation time distributions}\label{Sec:Distr:Occ_Visits}
In the remaining part of this section we will briefly present the results obtained in \cite{ROAP} regarding the distribution of the positive-axis occupation time and the distribution of returns number at the origin. We consider these quantities together because they can both be determined by the form of the generating function $F(z)$ (or equivalently by the form of $P(z)$) considered previously in this section.

The first quantity we deal with is the distribution up to $n$-th step of the positive-axis occupation time $K_n$, which indicates the number of steps spent by the particle in the set $x>0$. For instance, this quantity is studied in all that processes where one is interested in the fraction of time that a order parameter, for example the magnetization of a ferromagnet, has assumed positive values, see \cite{God-Luck, Bel-Barkai, Kor-Barkai}.

To find the asymptotic distribution of the positive-axis occupation time we make use of the Lamperti theorem \cite{Lamperti} that we will explain in the following. The theorem first requires that the stochastic process is recurrent in a state $\sigma$ and that the system can be divided into two subsets $A$ and $B$ that communicate each other through $\sigma$. More precisely we require that if the particle is in $A$ ($B$) at step $n-1$ and it is in $B$ ($A$) at the step $n+1$, then at step $n$ the particle needs to be in $\sigma$. In the Gillis process, which is symmetric with respect to the origin, the natural choice to divide the states is defining as the set $A$($B$) the positive(negative) axis, while obviously $\sigma=0$.  
Another request is the existence of the following limits
\begin{equation}
\lim_{n\to\infty}\mathbb{E}\left(\frac{K_n}{n} \right)=\eta
\label{Lamp:eta}
\end{equation}
and
\begin{equation}
\lim_{z\to 1^-} \frac{\left(1-z\right)F'(z)}{(1-F(z))}=\delta,
\label{Lamp:delta_1}
\end{equation}
where $F(z)$ is, as before, the generating function of the first return probability $F_n$ and the occupation time $K_n$ is defined with the convention that the occupation of the origin is counted or not according to whether the last occupied state was in $A$ or in $B$. The second limit is equivalent to the following requirement for the form of $F(z)$ \cite{Lamperti}:
\begin{equation}
F(z)=1-(1-z)^\delta L\left(\frac{1}{1-z}\right).
\label{Lamp:delta_2}
\end{equation}
If both the conditions are satisfied with $0\leq \eta \leq 1$ and $0\leq \delta \leq 1$, then
\begin{equation}
\lim_{n\to \infty} \mathrm{Pr}\left\lbrace \frac{K_n}{n}\leq u \right\rbrace=G_{\eta,\delta}(u)
\end{equation}
exists (the Lamperti distribution $G_{\eta,\delta}(u)$ will be defined soon).

Let us consider the Gillis model. First of all we notice, as we have underlined previously, that the process is recurrent for $\epsilon\geq-\frac{1}{2}$, thus the theorem can be applied only in this range. Regarding the parameters \eref{Lamp:eta} and \eref{Lamp:delta_1} we can state that, due to the symmetry of the model, we have $\eta=1/2$ and $\delta$ is straightforward since $F(z)$ \eref{F(z):form} is exactly the same of \eref{Lamp:delta_2}, therefore:
\begin{equation}
\delta=\cases{
0&for $\epsilon=-\frac{1}{2}$\\
\frac{1}{2}+\epsilon~~~&for $-\frac{1}{2}<\epsilon\leq\frac{1}{2}$\\
1&for $\frac{1}{2}<\epsilon<1$.\\
}
\label{delta:lamp}
\end{equation}
 At this point Lamperti theorem gives us the distribution of $u=K_n/n$ as $n\to\infty$ \cite{Lamperti}:
\begin{itemize}
\item for $\epsilon=-\frac{1}{2}$
\begin{equation}
G_{\eta,0}(u)=1-\eta=\frac{1}{2};
\label{Lamp:=-1/2}
\end{equation}
\item for $\epsilon \in \left(-\frac{1}{2},\frac{1}{2}\right)$ 
\begin{equation}
G'_{\eta,\delta}(u)=\frac{a \sin(\pi\delta)}{\pi}\frac{u^\delta(1-u)^{\delta-1}+u^{\delta-1} (1-u)^{\delta} }{a^2 u^{2\delta}+2au^\delta (1-u)^\delta\cos(\pi\delta)+(1-u)^{2\delta}},
\label{Lamp:<1/2}
\end{equation}
where 
\begin{equation}
a=\frac{1-\eta}{\eta}=1;
\end{equation}
\item for  $\epsilon \in\left(\frac{1}{2},1\right)$, we have
\begin{equation}
G_{\eta,1}(u)=\cases{
0&for $u<\eta$\\
1&for $u\geq\eta$.\\
}
\label{Lamp:>1/2}
\end{equation}
\end{itemize}

We observe that for $\delta=1$, which corresponds to $\epsilon\in\left( \frac{1}{2},1\right)$, we have that the distribution of $K_n/n$ is the Dirac delta function centered in $1/2$, meaning that the walker spends half of time in the positive axis and the other half in the negative one. This is a direct consequence of the existence of the stationary distribution in this interval; indeed, due to the Birkhoff's ergodic theorem one has that the time average of the occupation time $K_n$ of the set $A$ converges asymptotically to the ensemble average of the function
\begin{equation}
\theta(j)=\cases{
0& for $j<0$\\
\frac{1}{2}& for $j=0$\\
1& for $j>0$,
}
\end{equation}
 which is the sum of the characteristic function of the set $A=\lbrace j:j>0\rbrace$ and one half the Kroenecker delta of the origin (this contribution is due to the convention made previously about the origin that in average is counted only half the time). Now, by considering the symmetry with respect to the origin of the stationary solution $\pi_j$ (see Section \ref{Sec:Stationary:Solution}), one has that the ensemble average of $\theta(j)$ is equal to $1/2$. Therefore one has:
 \begin{equation}
 \lim_{n\to \infty} \frac{K_n}{n}=\frac{1}{2}
 \end{equation}
and, as a result, the distribution of $K_n/n$ must tend as $n$ becomes large to a Dirac delta centered in $1/2$.

\begin{figure*}[h!]
	\centering
	\begin{tabular}{c @{\quad} c }
		\includegraphics[width=.45\linewidth]{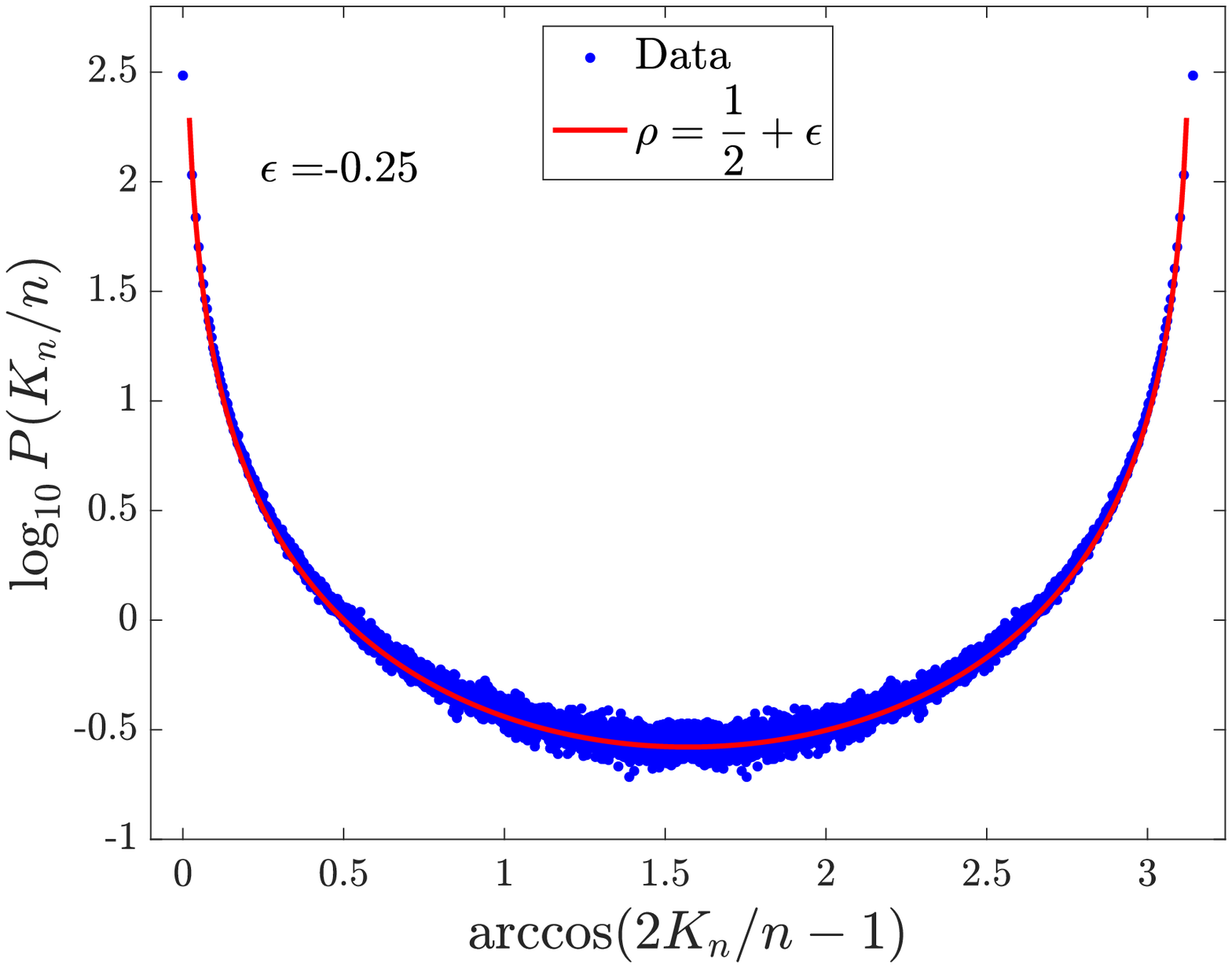} &
		\includegraphics[width=.45\linewidth]{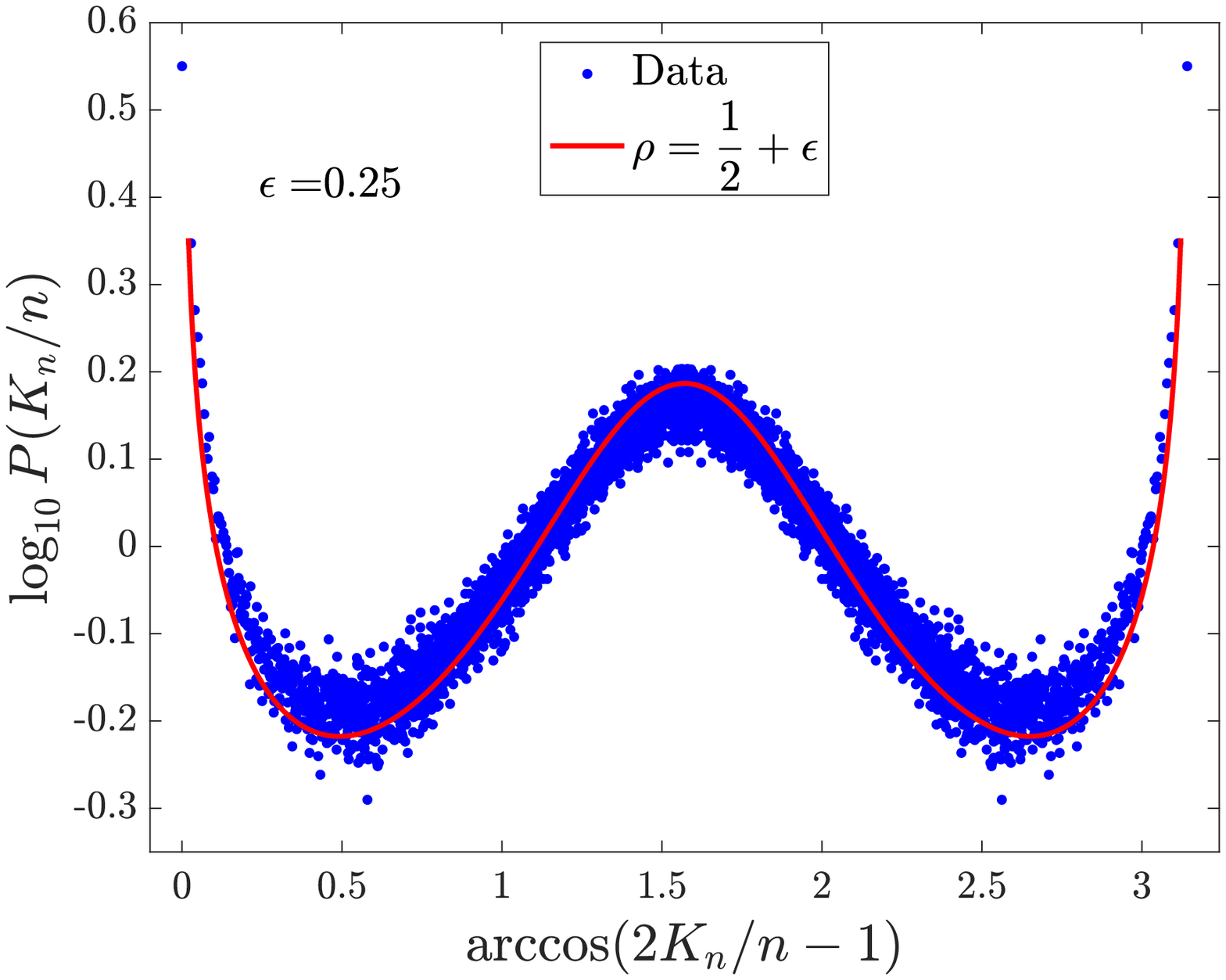} \\
		\small (a) & \small (b)
	\end{tabular}	
		\begin{tabular}{c @{\quad} c }
		\includegraphics[width=.45\linewidth]{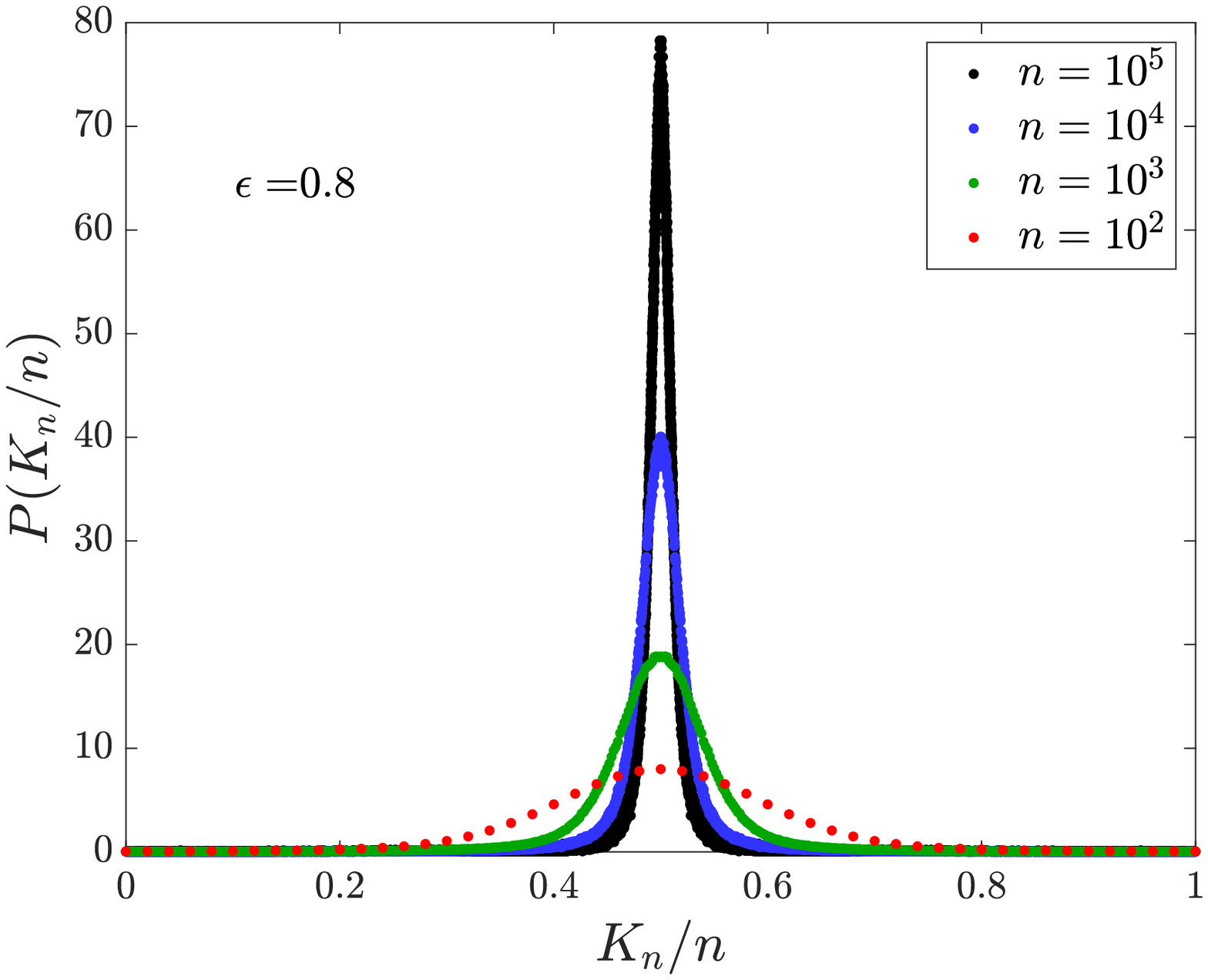} \\
		\small (c) 
	\end{tabular}	
	\caption{Distribution of the fraction of time $K_n/n$ spent by the particle in $A$ up to $n$ steps for the Gillis model in one dimension. On the top the (blue) dots present the distributions obtained for $\epsilon=-0.25$ (left) and $0.25$ (right) by simulating $10^6$ walks up to $n=10^4$, while the (red) line shows the Lamperti distributions with parameter $\delta$ given by \eref{Lamp:<1/2}. On the bottom the figure presents the distribution at different number of steps obtained by simulating $10^6$ walks with $\epsilon=0.8$. As the number of steps increases one observes that the distribution converges to a Dirac delta centered in $K_n/n=0.5$. }
\label{fig:1D:Lamp}
\end{figure*}

In fig.\ref{fig:1D:Lamp} we present the results obtained by simulations for the distribution of $K_n/n$ for three different values of $\epsilon$ that show the possible forms that the distribution may assume: for $\epsilon\in \left(-\frac{1}{2},0\right)$ one has a U-shaped distribution; while for $\epsilon$ from $0$ to $\frac{1}{2}$ the distribution assumes a W-shape; finally for $\frac{1}{2}\leq \epsilon <1$ one has, as we have already underlined, that the distribution is a Dirac delta. 

Now let us consider the distribution of the occupation time of the origin that, in our case, corresponds to the number of steps that end at $j=0$, therefore it is equivalent to the number of visits at the origin $V_n$. When one is interested in the distribution of these kind of variables it is possible to refer to a well known result by Darling and Kac \cite{Darling-Kac}, which states that for a Markov process the asymptotic distribution of the occupation time of a set of finite measure has the form of the Mittag-Leffler distribution
\begin{equation}
\mathcal{M}_\mu(\xi)=\frac{1}{\mu \xi^{1+\frac{1}{\mu}}}L_\mu \left(\frac{1}{\xi^\frac{1}{\mu}} \right),
\end{equation}
where $L_\mu(x)$ denotes the Lévy one-sided density of parameter $\mu$. 
Moreover, in \cite{ROAP} it has been considered the case in which the occupation set corresponds to the starting point and it has been proved that for renewal processes satisfying Lamperti theorem the parameter $\mu$ must be equal to the Lamperti parameter $\delta$, namely

\begin{equation}
\mu=\cases{
0&for $\epsilon=-\frac{1}{2}$\\
\frac{1}{2}+\epsilon&for $-\frac{1}{2}<\epsilon<\frac{1}{2}$\\
1&for $\frac{1}{2}\leq\epsilon<1$,
}
\label{1D:ML:parameter}
\end{equation}
and also that the variable $\xi$ is given by
\begin{equation}
\xi=\lim_{n\to\infty}\frac{1}{\Gamma(1+\mu)}\frac{V_n}{\langle V_n\rangle},
\end{equation}
with $\langle V_n\rangle$ denoting the average of $V_n$ over all walks and characterized by the asymptotic behaviour
\begin{equation}\label{Mean:visits}
\langle V_n\rangle \sim \frac{1}{\Gamma(1+\mu)}n^\mu H(n)
\end{equation}
where $H(n)$ is the slowly varying function characterizing the form of $P(z)$ in \eref{P(z):Form}.

We underline that the case $\mu=1$ is degenerate and one has the convergence
\begin{equation}\label{ML:deg}
\xi=\frac{1}{H(n)n}V_n\to 1
\end{equation}
 in probability, which is a sort of weak ergodic theorem \cite{Darling-Kac}. In fact, as we have seen in Section \ref{Sec:Stationary:Solution}, Birkhoff's theorem holds in the interval $\left(\frac{1}{2},1\right)$ and one has that the time average $V_n/n$ of the returns number must converge to the ensemble average of  $\delta_{0,j}$ that is given by $\pi_0$, i.e. the value of the stationary distribution at the origin. Moreover, in Section \ref{Sub_Sec:Pn} we have seen that the limiting value of $H(n)$ converges to a constant that is given by $\pi_0$; therefore, eqs. \eref{ML:deg} and \eref{Ret:Birkhoff} are equivalent and imply that the distribution of the number of visits at the origin rescaled by its mean value tends to a Dirac delta centered in $\xi=1$. 

\begin{figure*}[h!]
	\centering
	\begin{tabular}{c @{\quad} c }
		\includegraphics[width=.45\linewidth]{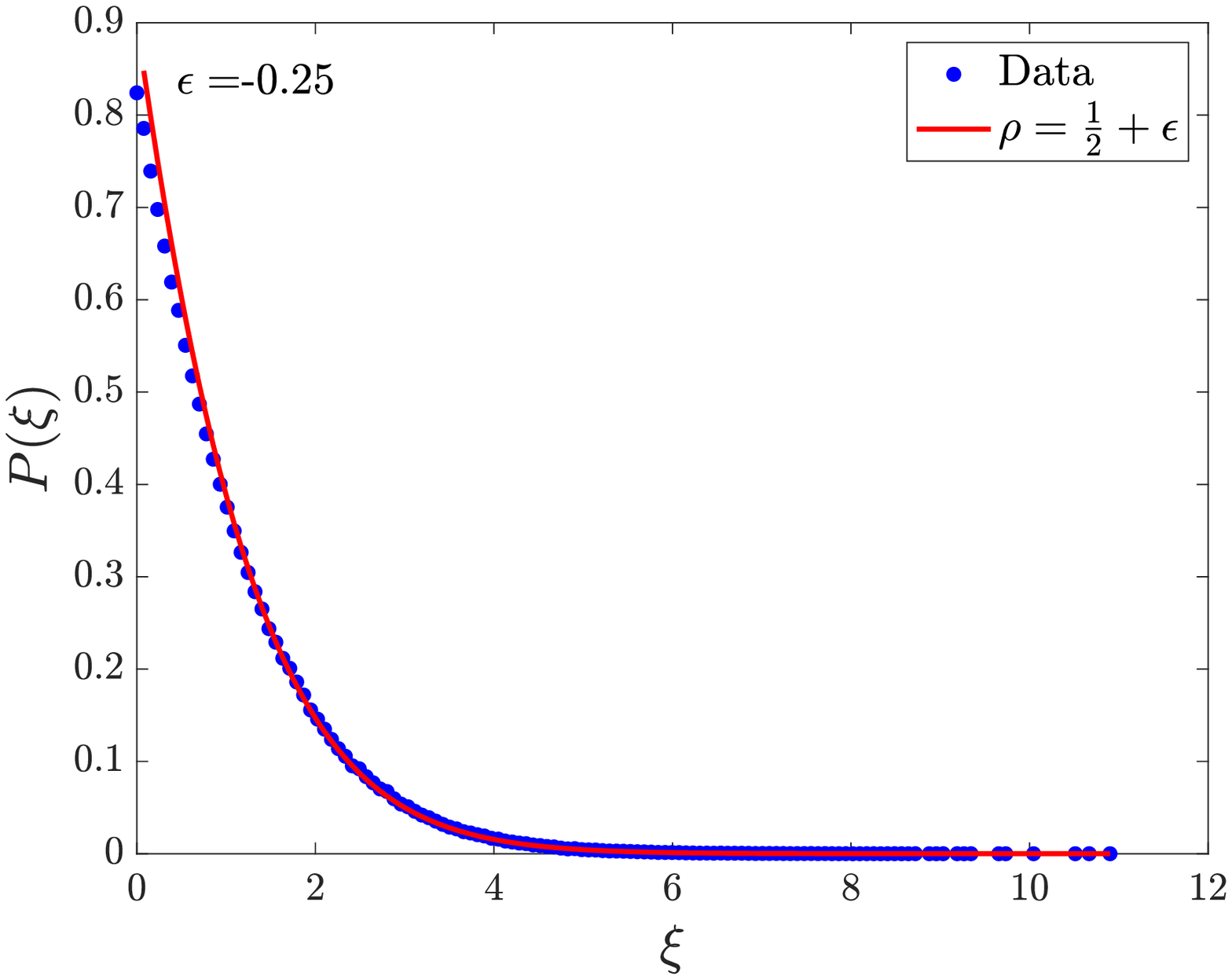} &
		\includegraphics[width=.45\linewidth]{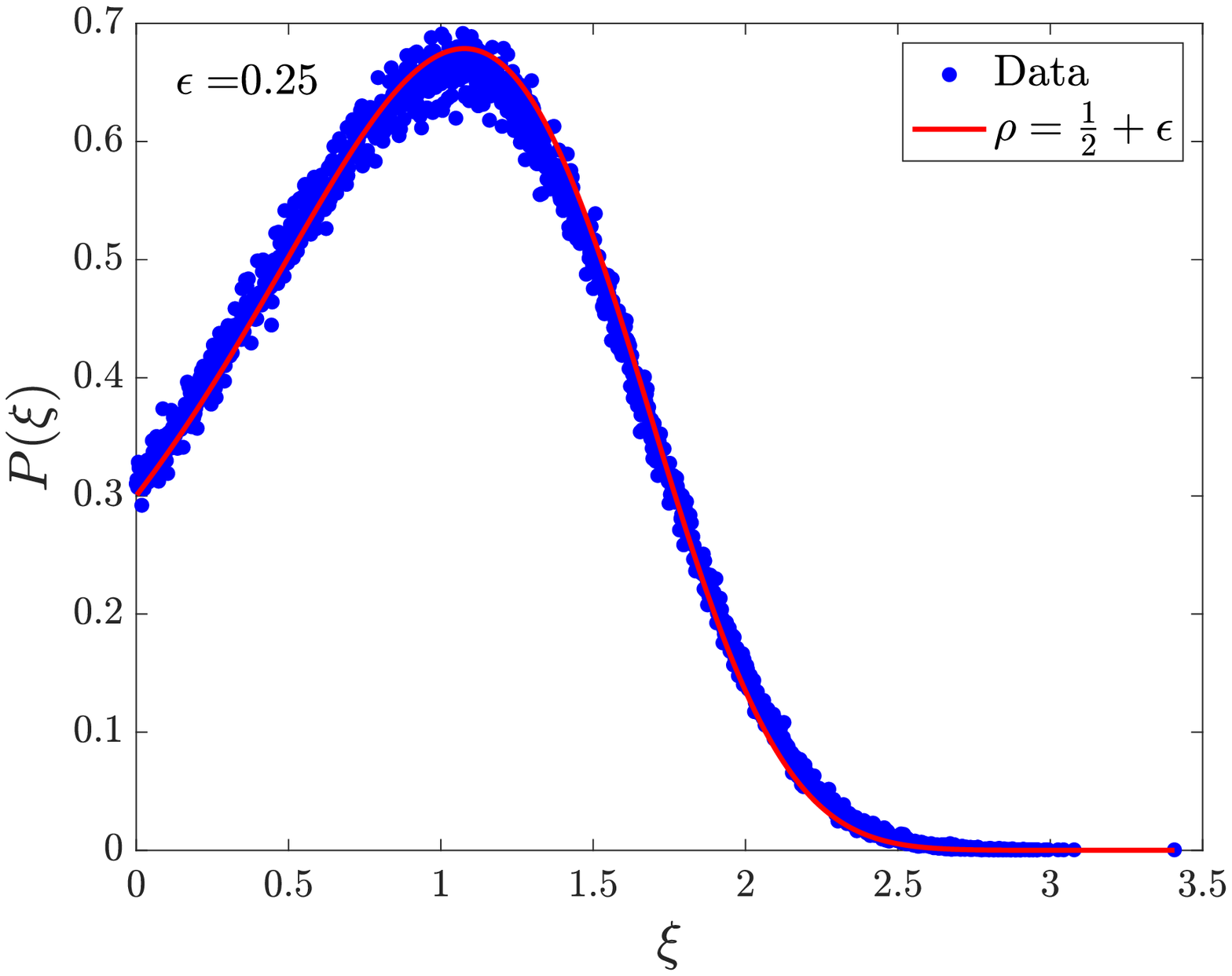} \\
		\small (a) & \small (b)
	\end{tabular}	
		\begin{tabular}{c @{\quad} c }
		\includegraphics[width=.45\linewidth]{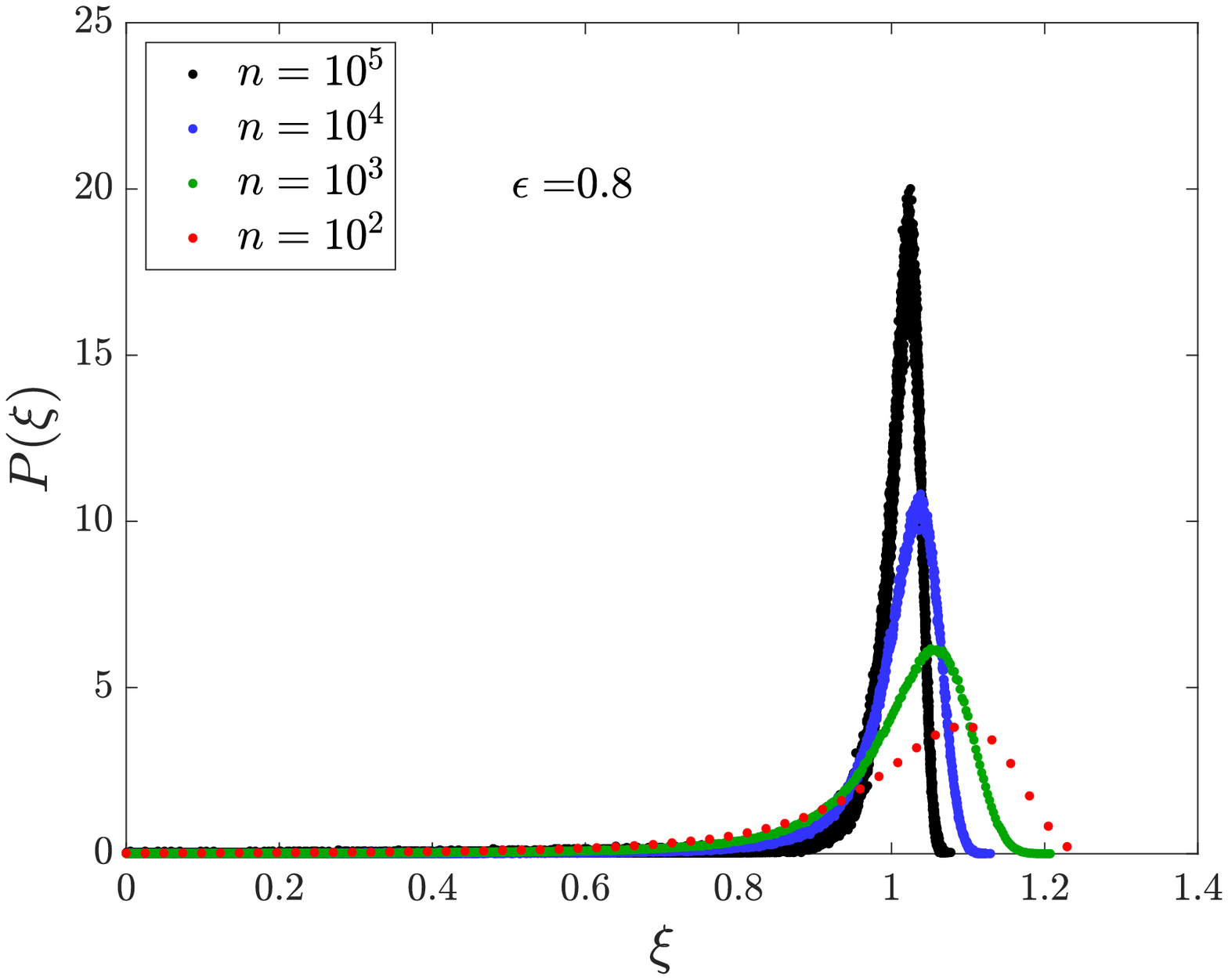} \\
		\small (c) 
	\end{tabular}	
	\caption{Distribution of the rescaled numbers of returns to the origin for the Gillis model in one dimension. On the top the figure presents the result (blue dots) obtained by simulating $10^6$ walks up to $10^4$ steps for $\epsilon=-0.25$ (left) and $\epsilon=0.25$ (right) and the Mittag-Leffler distribution (red line) with parameter $\mu$ given by \eref{1D:ML:parameter}. On the bottom the figure presents the distribution at different number of steps $n$ obtained by simulating $10^6$ walks with $\epsilon=0.8$. As the number of steps increases one observes that the distribution converges to a Dirac delta centered in $\xi=1$. }
\label{fig:1D:ML}
\end{figure*}

 In fig.\ref{fig:1D:ML} we show the simulations for values of $\epsilon$ that represent different behaviours of the Mittag-Leffler distribution: for $\epsilon\in\left[-\frac{1}{2},0\right]$ the distribution is monotonically decreasing and in the particular case $\epsilon=-\frac{1}{2}$ the decay is pure exponential; as $\epsilon$ increases from $0$ to $\frac{1}{2}$ the distribution has one maximum that gets closer to $\xi=1$; while for $1/2\leq \epsilon <1$ the distribution is a Dirac delta centered in $\xi=1$. 

\section{Continuum limit and transport properties}\label{Sec:3}
In this section we will consider an appropriate continuum limit leading to a diffusion equation for the probability density function (PDF) $p(x,t)$ of the process, through which we get the whole moments spectrum.

First of all let us consider the master equation that governs the evolution of the probability $P_n(j)$ of finding the particle at site $j$ after $n$ steps:
\begin{equation}
P_{n+1}(j)=P_{n}(j-1)\mathcal{R}(j-1)+P_{n}(j+1)\mathcal{L}(j+1),
\label{Pdf:master:eq}
\end{equation}
with initial condition
\begin{equation}
P_0(j)=\delta_{j,0},
\end{equation}
where the transition probabilities $\mathcal{R}(i)$ and $\mathcal{L}(i)$ are given by \eref{Gillis:1d_1} and \eref{Gillis:1d_2}.
Let $\delta x$ be the lattice spacing  and $\delta t$ the time step and define
\begin{equation}\label{Def:x:t}
x=j\delta x\quad\mathrm{and} \quad	t=n\delta t,
\end{equation}
while the probability density function $p(x,t)$ of being at position $x$ at time $t$ is related to $P_n(j)$ through
\begin{equation}\label{Def:p(x,t)}
P_n(j)=p(x,t)\delta x.
\end{equation}
Inserting \eref{Def:x:t} and \eref{Def:p(x,t)} in \eref{Pdf:master:eq}, one obtains
\begin{equation}
p(x,t+\delta t)=\frac{1}{2}\left(1-\frac{\epsilon \delta x}{x-\delta x} \right) p(x-\delta x,t) +\frac{1}{2}\left(1+\frac{\epsilon \delta x}{x+\delta x} \right) p(x+\delta x,t).
\end{equation}
Now let us expand the quantities in this relation up to the first order in $\delta t$ and the second order in $\delta x$ and consider the limit $\delta x,\delta t \to 0$ with the diffusion approximation, i.e. keeping constant the ratio $\delta x^2/\delta t=D_0$: in this way we get the Fokker-Planck equation
\begin{equation}
\frac{\partial p}{\partial t}=\frac{1}{2}\frac{\partial^2 p}{\partial x^2}+ \epsilon \frac{\partial }{\partial x}\left(\frac{1}{x}p\right),
\label{FP:Gillis}
\end{equation}
where we set $D_0=1$ due to the definition of the discrete model where both $\delta x$ and $\delta t$ are equal to $1$.
Equation \eref{FP:Gillis} describes the diffusion of a particle in the presence of a logarithmic  potential tuned by the parameter $\epsilon$. For a thermal system, $\epsilon$ is proportional to the ratio of the strength of the potential energy, $V(x)=V_0 \log|x|$, to the thermal energy $k_B T$, viz., $\epsilon=V_0/2k_B T$.

First of all we need to make some considerations about the potential $U(x)\equiv V(x)/k_B T$ and the Fokker-Plank equation; in fact \eref{FP:Gillis} presents a singularity in $x=0$ that in the discrete model is avoided by the definition of the transition probabilities in the origin, see eq.\eref{Gillis:1d_2}, where we have equal probability of making a step to the right or to the left so that the process in this single point is the same of a simple symmetric random walk. For this reason we impose that in the continuum limit the particle diffuses freely, i.e. $U(x)=0$, in a neighbourhood of $x=0$, which we set to be $\left(-a,a\right)$;  we will see later how the parameter $a$ can be determined. Therefore the regularized evolution equation becomes
\begin{equation}
\frac{\partial p}{\partial t}=\frac{1}{2} \frac{\partial }{\partial x} \left(\frac{\partial p }{\partial x}-  F(x) p \right),
\label{FP:reg:Gillis} 
\end{equation}
where $F(x)=-U'(x)$, with 
\begin{equation}
U(x)=\cases{ 
0 &for $|x|<a$\\
2\epsilon \log \left( \frac{|x|}{a} \right)  &for $|x|\geq a$.
}
\end{equation}
A detailed analysis of eq.\eref{FP:reg:Gillis} and the method used to obtain its solutions can be found in \cite{Dechant}, here we only give the main results. 

We immediately notice that for $\epsilon>1/2$ a stationary solution exists and it is given by
\begin{equation}
p_{st}(x)=\cases{
\mathcal{N}&for $|x|<a$\\
\mathcal{N} \left|\frac{x}{a}\right|^{-2\epsilon}  &for $|x|\geq a$,
}
\label{P:st}
\end{equation}
where $\mathcal{N}=(2\epsilon-1)/4a\epsilon$, while for $\epsilon<1/2$ it is not normalizable. This solution is valid in the limit $t\to \infty$ and represents the equilibrium state of the process. We observe that $p_{st}(x)$ presents the same power-law decay of the stationary distribution \eref{Gillis:ST} of the GRW.  

It is also possible to evaluate the time-dependent solution at $t$ large but finite, see \cite{Dechant}.   For $1/2<\epsilon<1$ one has in the region $|x|>a$
\begin{equation}
p(x,t)\sim\frac{a^{2\epsilon-1}}{2\epsilon\Gamma\left(\epsilon-\frac{1}{2}\right)}\Gamma\left(\epsilon+\frac{1}{2},\frac{x^2}{2t}\right)|x|^{-2\epsilon},
\label{P:t:ergodic}
\end{equation} 
where $\Gamma(\alpha,z)=\int_z^{+\infty} e^{-u} u^{\alpha-1} du$ is the incomplete Gamma function, while for $|x|<a$ the stationary solution is the dominating term. We observe that for  $t\gg x^2$  the solution \eref{P:t:ergodic} tends to the stationary one \eref{P:st}, thus the central part of the PDF does not depend on time and it decays as $|x|^{-2\epsilon}$; moreover, this region becomes larger and larger as $t$ increases. This behaviour is observed also in the discrete model, where the stationary distribution $\pi_j$ \eref{Gillis:ST} describes correctly the probability distribution $P_n(j)$ only in an interval around the origin and such an interval becomes larger as the number $n$ of steps increases. Therefore we expect that \eref{P:t:ergodic} describes the discrete model outside the central part at large but finite $n$.

In order to verify this statement and to test the validity of the continuum limit, instead of evaluating the distribution of the position $j$ of the walker after $n$ steps, we consider the distribution of the scaling variable $z=j/\sqrt{n}$. Thus, taking into account \eref{P:t:ergodic} and substituting $x=z\sqrt{t}$, we obtain the distribution of $z$ outside the central region
\begin{equation}\label{p(z,t)_ergodic}
p(z,t)\sim  f(z) t^{\frac{1}{2}-\epsilon},
\end{equation}
where 
\begin{equation}\label{f(z)_ergodic}
f(z)=\frac{a^{2\epsilon-1}}{2\epsilon\Gamma\left(\epsilon-\frac{1}{2}\right)}\Gamma\left(\epsilon+\frac{1}{2},\frac{z^2}{2}\right)|z|^{-2\epsilon}
\end{equation}
is the scaling function. We point out that in \cite{Dechant, Kess_Barkai} the distribution $p(z,t)$ of eq.\eref{p(z,t)_ergodic} is called \textit{Infinite Covariant Density} (ICD) due to the non-integrable singularity in $z=0$. The presence of such a singularity suggests that the scaling does not hold around $z=0$ for any finite time, and indeed we have already seen (fig.\ref{fig:St}) that in the central region the GRW is correctly described by the stationary distribution $\pi_j$, which is time-independent.

\begin{figure*}[h!]
	\centering
		\begin{tabular}{c @{\quad} c }
		\includegraphics[width=.45\linewidth]{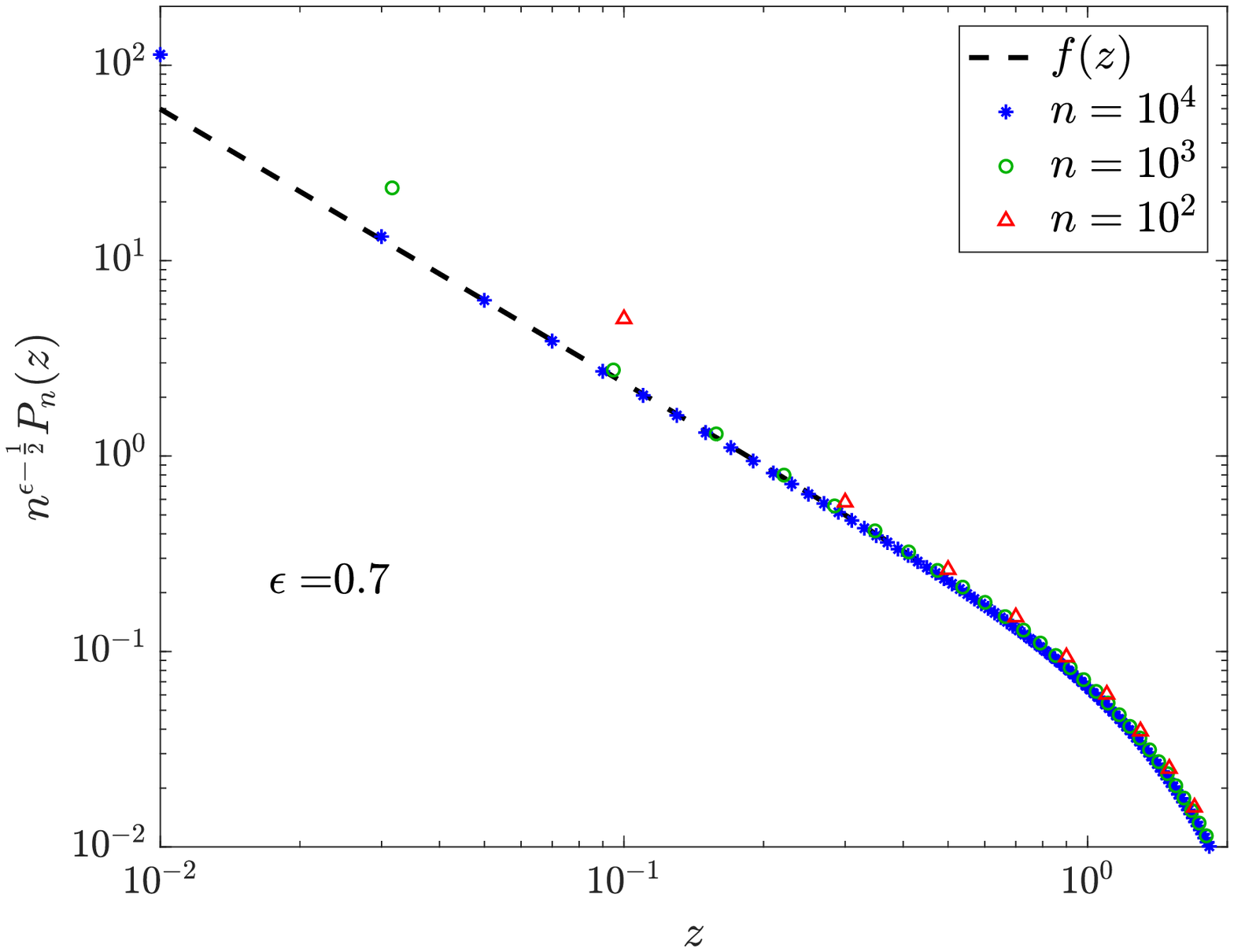} &
		\includegraphics[width=.45\linewidth]{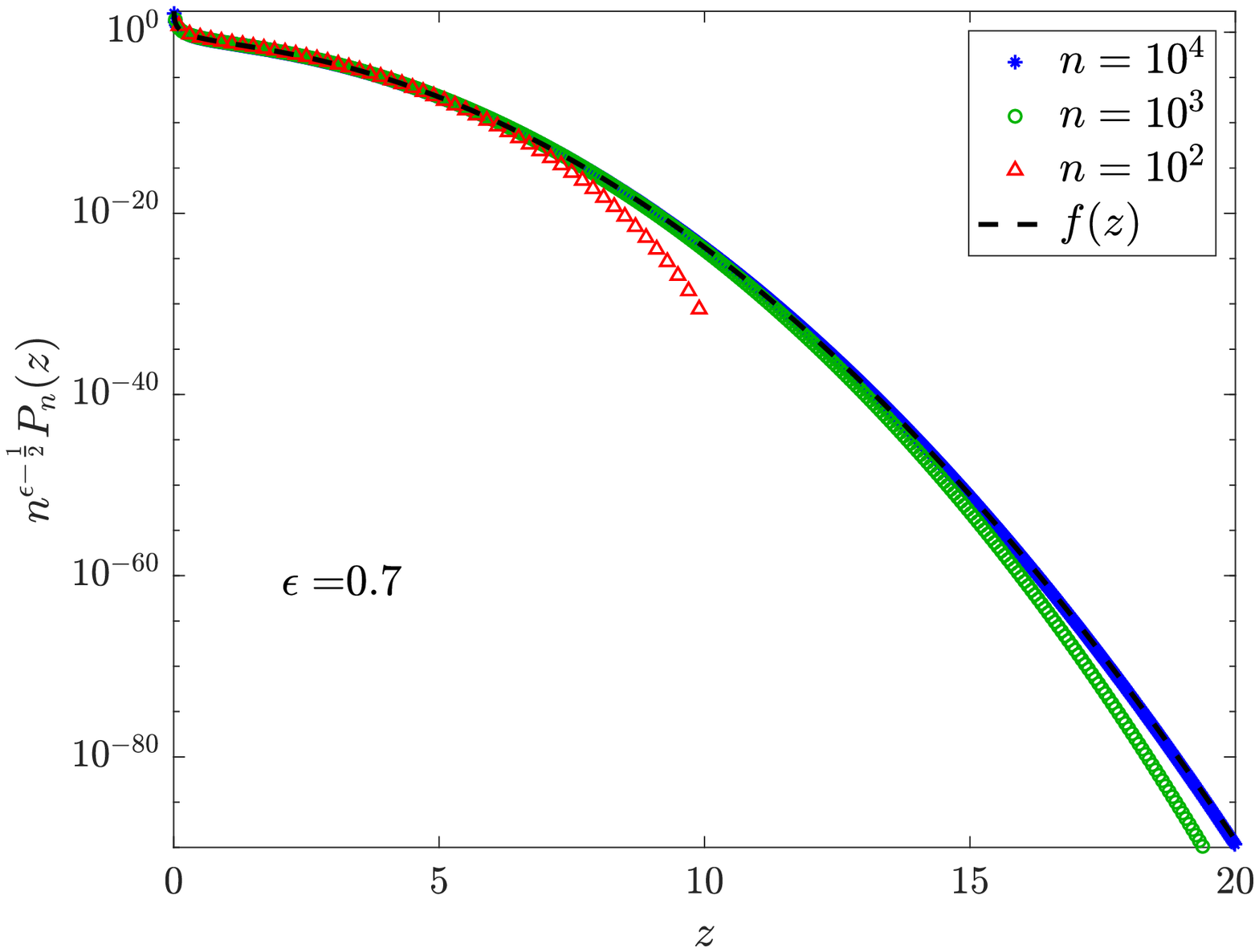} \\
		\small (a) & \small (b)
	\end{tabular}
	\begin{tabular}{c @{\quad} c }
		\includegraphics[width=.45\linewidth]{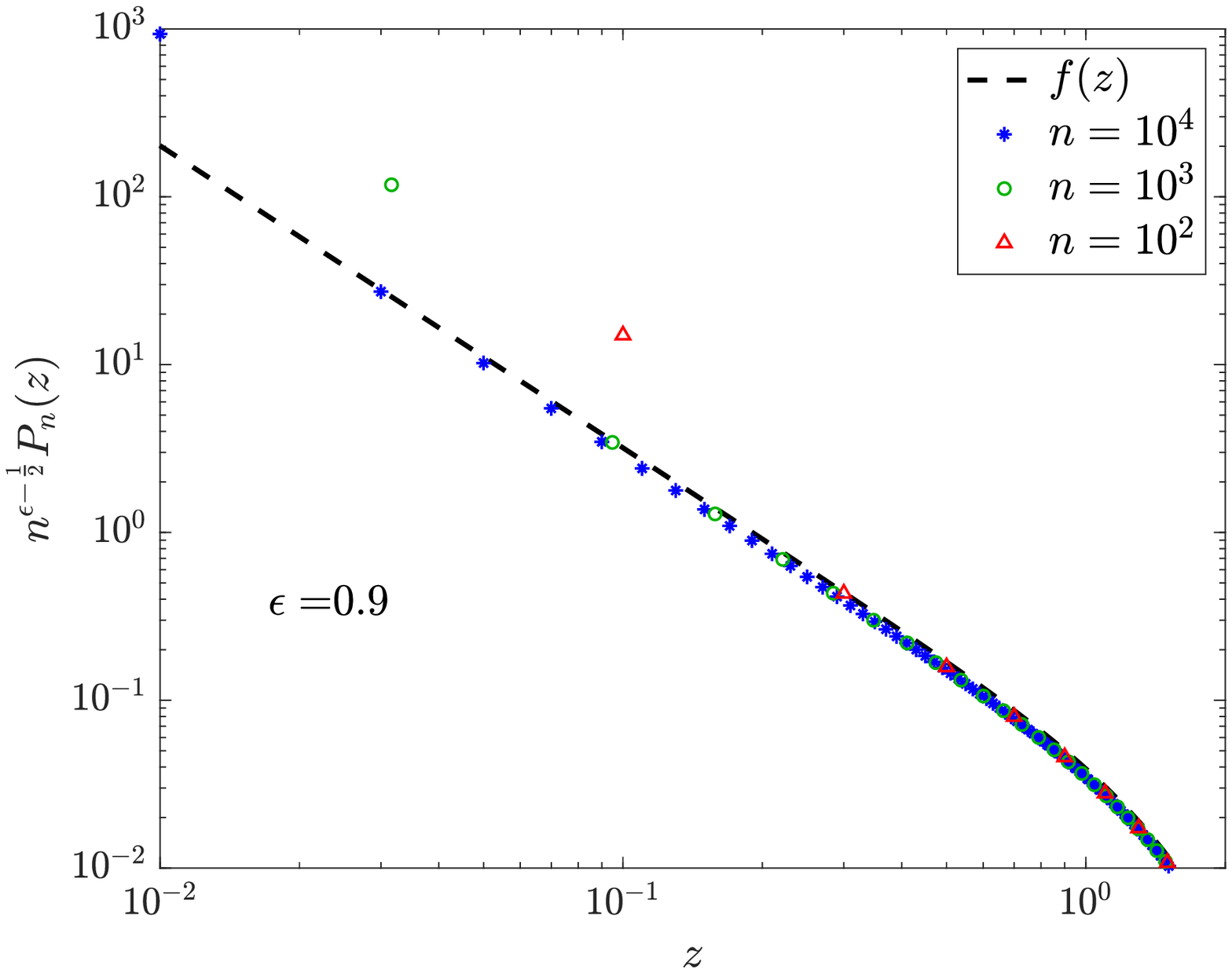} &
		\includegraphics[width=.45\linewidth]{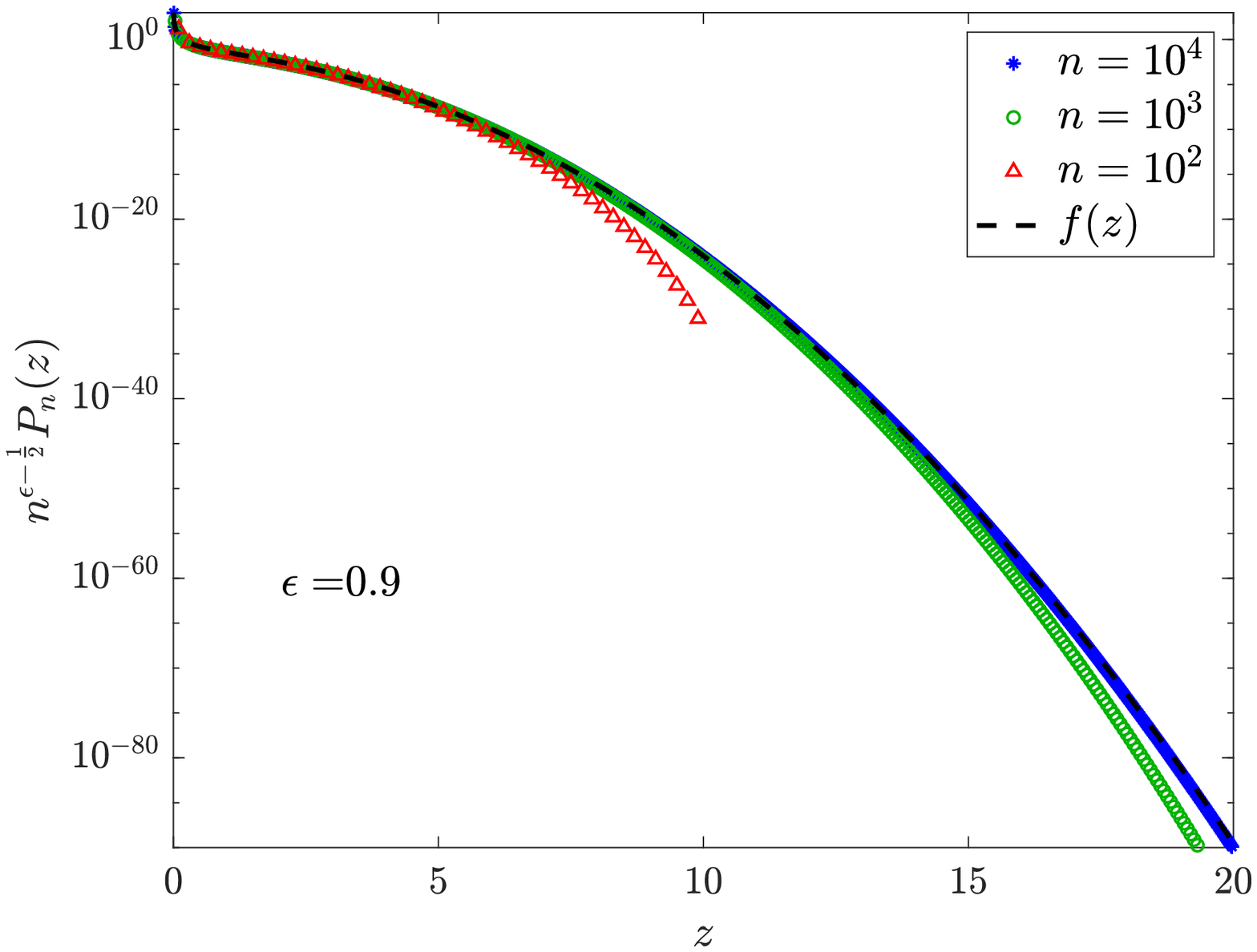} \\
		\small (c) & \small (d)
	\end{tabular}	
	\caption{Comparison between the distribution $P_n(z)$ of the variable $z=j/\sqrt{n}$, where $j$ is the position reached by the particle after $n$ steps, and the theoretical prediction.  The markers present the distribution $P_n(z)$ for different number of steps (triangles for $n=10^2$, circles for $n=10^3$ and stars for $n=10^4$), while the dashed lines correspond to  $f(z)$ given in \eref{f(z)_ergodic}. Figures (a) and (b), as well as figures (c) and (d), refer to the same case, presented with different scales: the figures on the left are in logarithmic scale, showing that the scaling does not hold for small z; those on the right are in semi-logarithmic scale on the y-axis, showing that the agreement is good outside the central region and for $n$ large enough.}
\label{fig:cont_vs_discr:ergodic}
\end{figure*}

In fig.\ref{fig:cont_vs_discr:ergodic} we observe that as the number of steps $n$ becomes larger eq.\eref{p(z,t)_ergodic} correctly predicts the distribution of the sacling variable at large $z$, i.e. at large distances from the origin. We remark that to get such an agreement, we have to tune the parameter $a$ in such a way that $p(z,t)$ can describe the discrete model. Therefore,  imposing that \eref{P:t:ergodic} at small $x$ reproduces the stationary distribution \eref{Gillis:ST}, we need to take $a$ such that 
\begin{equation}
\frac{a^{2\epsilon-1}}{2}=\frac{\Gamma(1+\epsilon)}{\Gamma(1-\epsilon)}.
\end{equation}

For $-1<\epsilon<1/2$ the stationary solution of \eref{FP:reg:Gillis} is not normalizable; therefore we only have the time-dependent solution which at large but finite time is evaluated as \cite{Dechant}:
\begin{equation}
p(x,t)\sim\cases{
\frac{2^{\epsilon-\frac{1}{2}}}{a\Gamma\left(\frac{1}{2}-\epsilon\right)}\left(\frac{a^2}{t}\right)^{1/2-\epsilon} &for $|x|<a$\\
\frac{2^{\epsilon-\frac{1}{2}}}{\Gamma\left(\frac{1}{2}-\epsilon\right)} \frac{|x|^{-2\epsilon}}{t^{1/2-\epsilon}}e^{-\frac{x^2}{2t}}& for $|x|\geq a$.
}
\label{P:t:diffusive}
\end{equation}
This solution, except for the part in the regularized region, is the same obtained in \cite{Lamperti-1962} (see Theorem 2.1) for the GRW. Moreover, we underline that the central part decays with time as $t^{-(1/2-\epsilon)}$, which is the same decay of the probability of being at the origin for the discrete model, see eq.\eref{P_n:asympt}.

Now we consider again the scaling variable $z=x/\sqrt{t}$; in this case the distribution of $z$ is given by
\begin{equation}\label{p(z,t)-non:ergodic}
p(z,t)\sim\cases{
\frac{2^{\epsilon-\frac{1}{2}}}{\Gamma\left(\frac{1}{2}-\epsilon\right)}\left(\frac{a^2}{t}\right)^{-\epsilon} &for $|z|<\frac{a}{\sqrt{t}}$\\
g(z)& for $|z|\geq \frac{a}{\sqrt{t}}$,
}
\end{equation}
where
\begin{equation}\label{g(z)}
g(z)=\frac{2^{\epsilon-\frac{1}{2}}}{\Gamma\left(\frac{1}{2}-\epsilon\right)} |z|^{-2\epsilon}e^{-\frac{z^2}{2}},
\end{equation}
thus outside the central region $\left(-\frac{a}{\sqrt{t}},\frac{a}{\sqrt{t}}\right)$ the distribution $p(z,t)$ does not depend on $t$ explicitly. We observe that for positive values of $\epsilon$ the scaling function $g(z)$ present a singularity in $z=0$, which in this case is integrable. Another difference with the function $f(z)$ is that  $g(z)$ is independent of the parameter $a$, hence the system becomes insensitive to the size of the regularizing region in the long-time limit.

\begin{figure*}[h!]
	\centering
	\begin{tabular}{c @{\quad} c }
		\includegraphics[width=.45\linewidth]{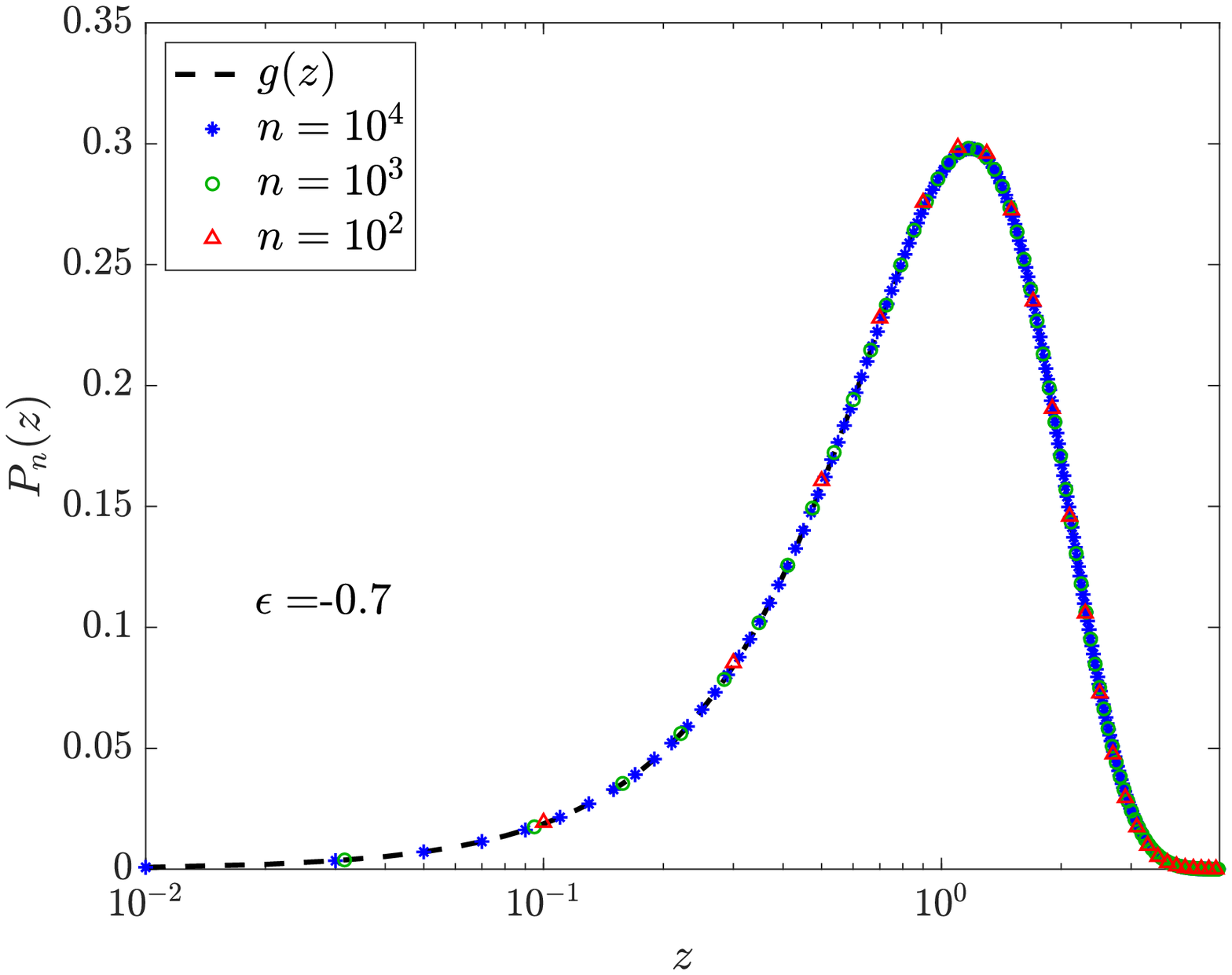} &
		\includegraphics[width=.45\linewidth]{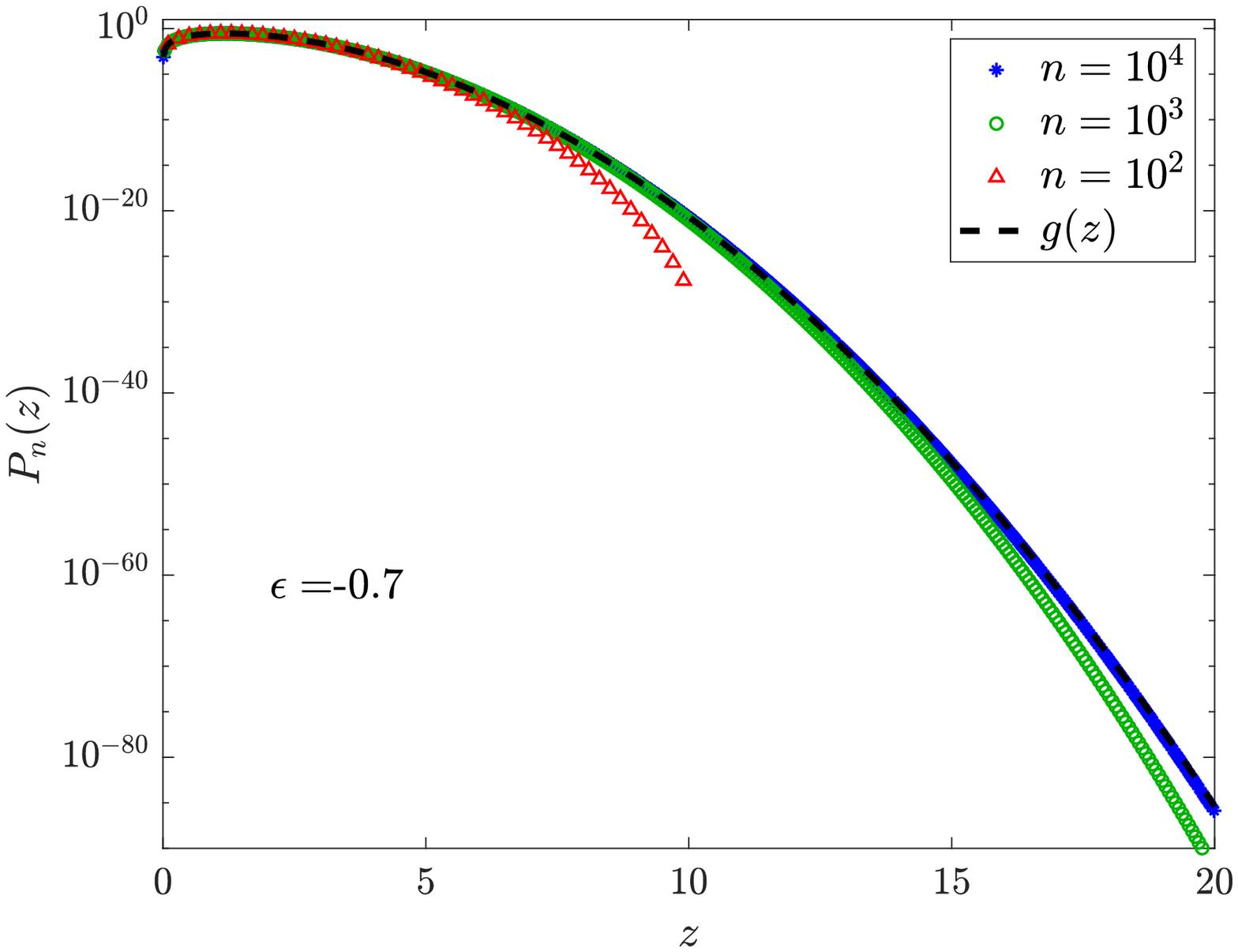} \\
		\small (a) & \small (b)
	\end{tabular}	
		\begin{tabular}{c @{\quad} c }
		\includegraphics[width=.45\linewidth]{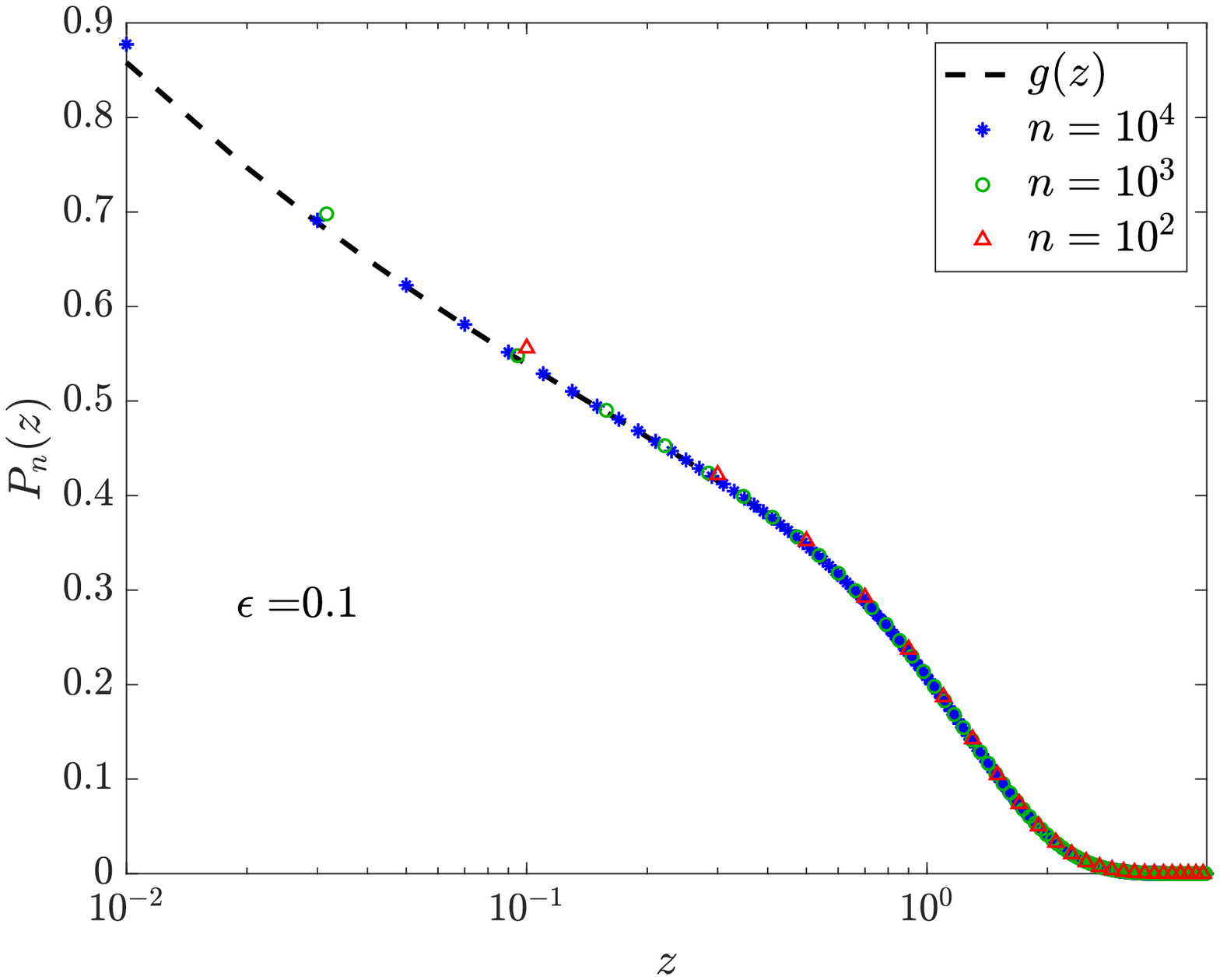} &
		\includegraphics[width=.45\linewidth]{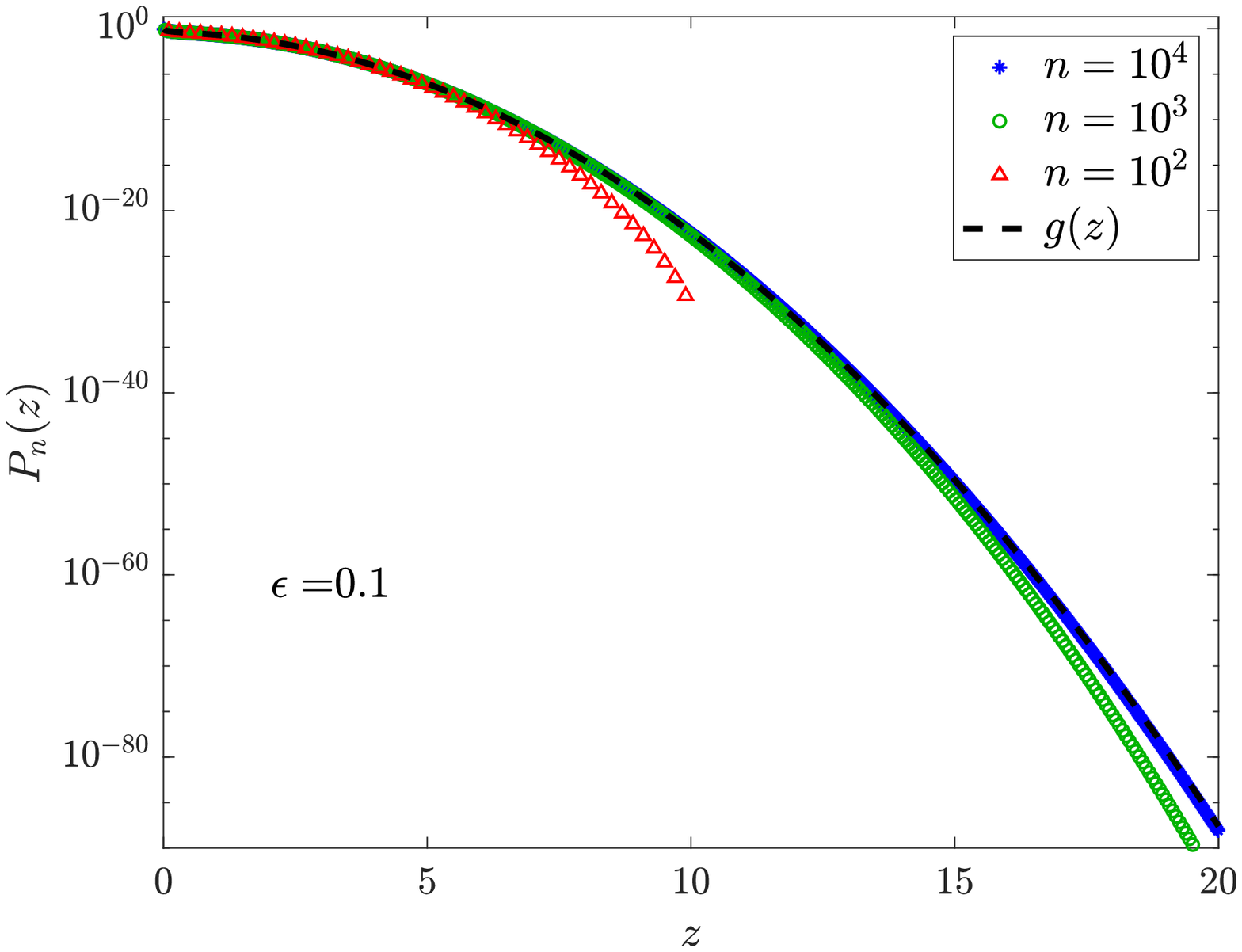} \\
		\small (c) & \small (d)
	\end{tabular}	
	\caption{Comparison between the distribution $P_n(z)$ of the variable $z=j/\sqrt{n}$, where $j$ is the position reached by the particle after $n$ steps, and the theoretical prediction.  The markers present the distribution $P_n(z)$ for different number of steps (triangles for $n=10^2$, circles for $n=10^3$ and stars for $n=10^4$), while the dashed lines correspond to  $g(z)$ given in \eref{g(z)}. Figures (a) and (b), as well as figures (c) and (d), refer to the same case, presented with different scales: the figures on the left are in semi-logarithmic scale on the x-axis; those on the right are in semi-logarithmic scale on the y-axis. The two different scales are used to show the agreement between numerical results and theoretical ones both at small and at large $z$ for $n$ large enough.}
\label{fig:cont_vs_discr}
\end{figure*}

In fig.\ref{fig:cont_vs_discr} we observe that as the number of steps increases the distribution of $z$ collapses to the scaling function $g(z)$. This behaviour is valid both for small $z$, see figures \ref{fig:cont_vs_discr}(a) and \ref{fig:cont_vs_discr}(c), and for large $z$, figures \ref{fig:cont_vs_discr}(b) and \ref{fig:cont_vs_discr}(d).

Through the continuum limit we have demonstrated that the GRW is closely related to a diffusing particle in the presence of a logarithmic potential, consequently we expect that also the transport properties of the two systems are related. Therefore we will use the probability density function $p(x,t)$ to obtain the moments spectrum of the discrete model.

For $-1<\epsilon<\frac{1}{2}$ we have seen that outside the central region the distribution of the scaling variable $z$ is described by $g(z)$. This function is characterized by an integrable singularity in $z=0$ and therefore all the moments $\langle|z|^q\rangle$ of $g(z)$ exist and are constant. Consequently all the moments of $p(x,t)$ scale like normal diffusion, namely
\begin{equation}
\langle |x|^q\rangle_t \sim t^{q/2}.
\label{spectrum:normal}
\end{equation}
We underline that the whole moments spectrum is governed by the same scaling $x\sim t^{1/2}$ characterizing the probability density function of eq.\eref{p(z,t)-non:ergodic} outside the central part; on the contrary, the PDF at the origin decays with a different scaling, namely $p(0,t)\sim t^{-\frac{1}{2}+\epsilon}$ given in eqs.\eref{P_n:asympt} and \eref{p(z,t)-non:ergodic}, and this represents an exception with respect to the standard relation between the scaling of the moments spectrum, the scaling of the PDF and the return probability \cite{Orbach}.

Contrary to the case above, the interval $\frac{1}{2}<\epsilon<1$ is characterized by the existence of a stationary distribution $p_{st}(x)$ that is reached in the long time limit, but, due to its slow decay $|x|^{-2\epsilon}$, only the low-order ($q<2\epsilon-1$) moments of $p_{st}(x)$ are finite, while the high-order ones diverge. To obtain the time-dependence of these moments it is possible to use the infinite covariant density $p(z,t)$ of eq.\eref{p(z,t)_ergodic}, see \cite{Dechant, Kess_Barkai}, which presents a non-integrable singularity in $z=0$, where the scaling does not hold. Due to the presence of such a singularity, the low-order moments are not measurable with respect to the ICD, but the high-order moments ($q>2\epsilon-1$) are finite and one has:
\begin{equation}
\langle |x|^q \rangle_t \sim t^{\frac{1+q}{2}-\epsilon}\cdot 2\int_{0}^{\infty} |z|^q f(z) dz.
\end{equation}
To summarize, the whole moments spectrum for $\epsilon>\frac{1}{2}$ is given by
\begin{equation}
\langle | x |^q \rangle_t \sim \cases{
\mathcal{K} &for $q<2	\epsilon-1$\\
t^{\frac{1+q}{2}-\epsilon} &for $q>2	\epsilon-1$.
}
\label{spectrum:ergodic}
\end{equation}
Therefore the model is strongly anomalous, see \cite{Cast}, in the ergodic regime with the second moment increasing slower than linearly, namely $\langle | x |^q \rangle_t\sim t^{\frac{3}{2}-\epsilon}$. We point out that in this system the strong anomalous diffusion arises from the fact that the low-order moments are yielded by the stationary distribution, whereas the high-order moments are computed through the Infinite Covariant Density. A different mechanism leading to strong anomalous diffusion regards the occurrence of rare events governing the dynamics, as described by the Big Jump Principle, see \cite{Vez_Bar_Bur:2019, Vez_Bar_Bur:2020, Bur_Vez}.

\begin{figure}[h!!!]
        \centering
        \includegraphics[scale=0.4]{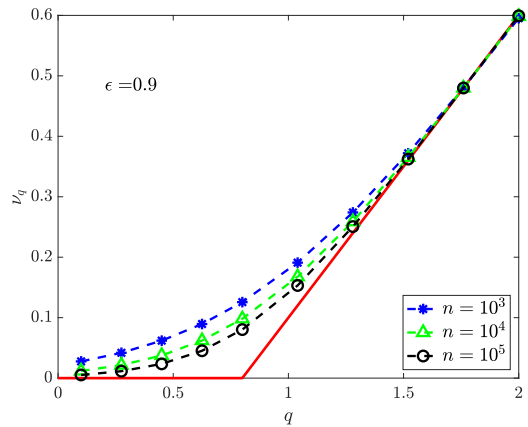}
\caption{Exponent $\nu_q$ characterizing the asymptotic power-law growth of the $q$-th moment for the Gillis random walk with $\epsilon=0.9$. Data are obtained by considering different total number $n$ of steps: asterisks correspond to $n=10^3$, triangles to $n=10^4$ and circles to $10^5$. The (red) line is the theoretical prediction \eref{spectrum:ergodic}.}\label{fig:1D:Spec:Erg}
\end{figure}

In fig.\ref{fig:1D:Spec:Erg} we present the moments spectrum in the anomalous regime for $\epsilon=0.9$. One easily observes the two different behaviours of exponent $\nu_q$ characterizing the power-law growth $t^{\nu_q}$ of the $q$-th moment: for $q<2\epsilon-1$ the moment $\langle | x |^q \rangle_t$ tends to a costant, while for $q>2\epsilon-1$ one has $\langle | x |^q \rangle_t\sim t^{\frac{1+q}{2}-\epsilon}$.

\section{Statistics of records and maximum}\label{Sec:4}
In recent years the statistics of records have attracted wide interest due to its applications in a large variety of fields, such as meteorology \cite{Bassett ,Benestad, Red-Peter}, hydrology  \cite{Vog-Zaf-Mat}, finance \cite{Sabir-Santh, Barlevy} and sports \cite{Gemb-Tay-Suter, Ben-Red-Vaz}: reviews \cite{Majumdar:FirstPassage, God-Maj-Sch} discuss in detail record statistics for stochastic processes in one dimension.

In this section we will deal with the statistics of the number of records and the statistics of the maximum. First of all, let us define the quantities of interest: given a sequence $\left\{X_0,X_1,\cdots,X_n\right\}$ of $n$ events, the event $X_i$ is called a record if its value exceeds all the previous data; while the maximum $M_n$ is naturally defined as the the biggest value of the entire sequence. In our case, since the motion occurs on the integer lattice with steps only between first neighbour sites and the starting site is the origin, i.e. $X_0=0$, which by definition is the first record, one has that the number of record $\mathcal{N}_n$ after $n$ steps satisfies
\begin{equation}
\mathcal{N}_n=M_n+1,
\end{equation}
therefore it is sufficient to study the distribution of the maximum $M_n$. To this aim we proceed as follows: firstly we divide the walk $\left\{X_0,X_1,\cdots,X_n\right\}$ up to the $n$-th step in shorter walks, called excursions, of which we can easily obtain the statistics of maximum; afterwards, starting from the knowledge of the properties of a single excursion, we obtain the behaviour of the expected maximum of the entire walk after $n$ steps.

For simplicity, due to the symmetry of the model, we will consider the Gillis random walk only on the positive integer axis, where the origin is considered as reflecting. An excursion is a subsequence $\left\lbrace X_0,\cdots,X_\tau \right\rbrace$ of the complete walk having the property that $X_\tau$ corresponds to the first return to the starting site $X_0$, namely $\tau$ is given by
\begin{equation}
\tau=\min\left\lbrace i>0 | X_i=0 	\right\rbrace.
\end{equation}
As we have seen previously, we have certain return to the starting site only for $\epsilon>-\frac{1}{2}$, then, in the following, we consider only this case. At this point we are interested in the maximum $m$ of an excursion, which is given by
\begin{equation}
m=\max_{0\leq j \leq \tau} X_j. 
\end{equation}
Obviously $m$ is a random variable and it can be characterized by the distribution $\mathcal{M}_{m}(x)=\mathbb{P}\left\{ m\leq x \right\}$, which is associated with another quantity characterizing the motion during an excursion. In fact, the request that the maximum of an excursion is smaller than a certain value $x$ corresponds to the request that the particle does not reach the position $x$ before the end of the excursion. Therefore, $\mathcal{M}_{m}(x)$ is equivalent to $1-Q(x)$, where $Q(x)$ is the probability of reaching $x$ before returning to the origin. To compute this quantity, we make use once again of the results of the continuum limit: indeed for a diffusing particle subject to a potential $V(x)$ it has been proved that $Q(x)$ is related to the potential by, see \cite{Maj-Rosso-Zoia}:
\begin{equation}
Q(x)\sim\left(\int_0^x e^{2V(x')}dx'\right)^{-1}.
\end{equation}
Taking into account the potential $V(x)=\epsilon\ln|x|$ obtained with the continuum limit, see Section \ref{Sec:3}, one has:
\begin{equation}\label{F_M:form}
\mathcal{M}_m(x)=1-Q(x)= 1- C x^{-\gamma},
\end{equation}
where $\gamma=2\epsilon+1$.
We obtained this result regarding the form of $\mathcal{M}_m(x)$ in a heuristic way, but the same can be obtained in a more formal way by the so called Lyapunov functions, which are deeply treated in several work about non-homogeneous random walk \cite{Lamperti-1960, Lamperti-1963, Hry-Men-Wade, Mens-Pop-Wade}. In \ref{App_D} we make use of these techniques to illustrate another method to obtain \eref{F_M:form}. 

Now let us consider the complete walk up to $n$ steps and suppose that it is composed by $N_n$ excursions. The distribution of the maximum $M_n$ of the entire walk is defined as
\begin{equation}
Q_n(x)=\mathbb{P}(M_n<x),
\end{equation}
and, knowing that in a walk of $n$ steps there are $N_n$ excursions, one can write
\begin{equation}
Q_n(x)=\mathcal{M}_m(x)^{N_n}=(1-Cx^{-\gamma})^{N_n}.
\end{equation}
To find the limiting distribution of $Q_n(x)$, let us consider the transformation $x=a_n z$ with,
\begin{equation}
a_n=(C N_n)^{\frac{1}{\gamma}}
\end{equation}
and take the limit
\begin{equation}
\lim_{n\to \infty} Q_n(a_n z)=\lim_{N_n\to\infty}\left(1- \frac{z^{-\gamma}}{N_n}\right)^{N_n}=e^{-z^{-\gamma}},
\end{equation}
where in the first equality we used the fact that in a recurrent process the number of visits at the starting site goes to infinity as $n\to\infty$. Therefore, the limiting distribution of the rescaled maximum
\begin{equation}\label{Maximum:Scaling}
\varepsilon=\frac{M_n}{(CN_n)^{\frac{1}{\gamma}}}
\end{equation}
is a Fr\'echet distribution, namely
\begin{equation}
\mathcal{Q}(z)=\lim_{n\to\infty}\mathbb{P}\left(\varepsilon<z\right)=e^{-z^{-\gamma}}.
\end{equation}
Consequently, due to the scaling form \eref{Maximum:Scaling} of the maximum, we have that $M_n\sim N_n^{1/\gamma}$.

\begin{figure}[h!!!]
        \centering
        \includegraphics[scale=0.4]{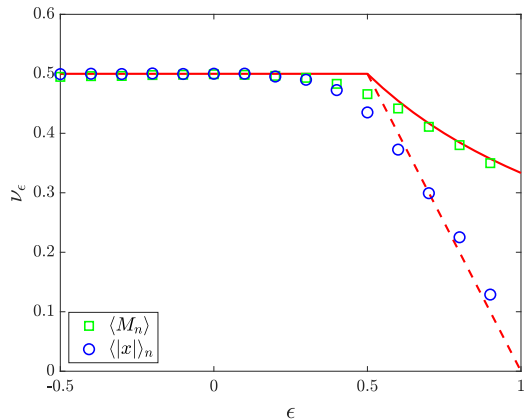}
\caption{Exponent $\nu_\epsilon$ characterizing the asymptotic power-law growth of the expected maximum $\langle M_n\rangle$ and the absolute first moment $\langle |x| \rangle_n$ depending on $\epsilon$. Data are obtained by considering $n=10^5$ numbers of steps and $10^6$ walks. The (green) squares are the exponents regarding the expected maximum, while the (blue) circles depict the exponents of the absolute first moment. The continuous (red) line refers to the exponent of $\langle M_n\rangle$ according to \eref{Asymp:Maximum}, while the dashed (red) line refers to the exponents of $\langle |x| \rangle_n$ according to \eref{spectrum:normal} and \eref{spectrum:ergodic}.}\label{Fig:1D:Records}
\end{figure}

Lastly we observe that the number of excursions $N_n$ that compose the random walk and the number of visits $V_n$ at the starting point up to time $n$ are related by $N_n=V_n-1$, due to the fact that the first visit corresponds to the beginning of motion, while the first excursion ends with the first return to the origin. Thus the mean number of excursions increases as, see eq.\eref{Mean:visits}:
\begin{equation}\label{Mean:excursions}
\langle N_n \rangle\sim n^\mu,
\end{equation}
from which we get the expected maximum after $n$ steps
\begin{equation}\label{Asymp:Maximum}
\langle M_n \rangle\sim n^\frac{\mu}{\gamma}=\cases{
n^\frac{1}{2} & $-\frac{1}{2}<\epsilon<-\frac{1}{2}$ \\
n^\frac{1}{1+2\epsilon} & $\epsilon\geq \frac{1}{2}$.
}
\end{equation}
We underline that for the ergodic regime, i.e. in $\left(\frac{1}{2},1\right)$, one has that the asymptotic behaviour of the expected maximum, or equivalently the mean number of records, is different from that of the absolute first moment given in eq.\eref{spectrum:ergodic}. This fact, that can be easily observed in fig.\ref{Fig:1D:Records}, contrasts with the result obtained in the interval $\left(-\frac{1}{2},\frac{1}{2}\right)$ or in other stochastic processes where the two quantities $\langle M_n \rangle$ and $\langle |x| \rangle_n$ have the same asymptotic growth, see for example \cite{Com-Maj, ROAC}.

\section{Generalization of the Gillis model}\label{Sec:5}
In this section we will introduce a generalization of the Gillis model, which we prove to be the discrete version of a diffusing particle subject to a force characterized by a power-law dependence on the distance from the origin. This kind of physical system is considered for instance in \cite{Bray}, regarding non-universal persistence and vortex dynamics, and in \cite{Agh_Kes_Bar_19, Agh_Kes_Bar_20}, dealing with infinite ergodic theory.

The model is defined through the probabilities $\mathcal{R}_\beta(j)$ and $\mathcal{L}_\beta(j)$ that the particle at the site $j\neq0$ makes a step to the right or to the left:
\begin{equation}\label{Gilli:Gen:1}
\mathcal{R}_\beta(j)=\frac{1}{2}\left(1-\textrm{sgn}(j)\frac{\epsilon}{|j|^\beta} \right) \quad \textrm{and} \quad \mathcal{L}_\beta(j)=\frac{1}{2}\left(1+ \textrm{sgn}(j)\frac{\epsilon}{|j|^\beta}\right),
\end{equation}
while if $j=0$
\begin{equation}\label{Gilli:Gen:2}
\mathcal{R}_\beta(j)=\mathcal{L}_\beta(j)=\frac{1}{2},
\end{equation}
where $\epsilon\in (-1,1)$  and $\beta>0$. The Gillis parameter $\epsilon$, as before, tunes the bias of the process: for positive values of $\epsilon$ the particle tends to move towards the origin, while for negative ones the particle escapes from it, the case $\epsilon=0$ corresponds to the simple symmetric random walk. The new parameter $\beta$ controls  the non-homogeneity along the lattice: for $\beta>1$ the bias due to $\epsilon$ decreases faster than the original model, which is recovered for $\beta=1$, while for $\beta<1$ the decrease is slower. 

To study the model let us start considering the moments of the increment with the aim to obtain some asymptotic properties by using Lamperti criteria, see section \ref{Sec:Lamperti:Criteria}. Firstly, as we have seen before, we need to restrict the stochastic process on the positive axis, but, due to the symmetry of the model, this requirement does not loose generality. Secondly, we only need to evaluate the first two increment moments. We have:
\begin{equation}
\mu_1(x)=-\frac{\epsilon}{x^\beta} \nonumber\quad\textrm{and}\quad \mu_2(x)=1.
\end{equation}
At this point we need to distinguish two cases according to the values of $\beta$.

For $\beta<1$, according to theorems  \ref{Th:Lamp:Rec/Trans} and \ref{Th:Lamp:Finite}, we have that the process is transient for $\epsilon<0$, while for $\epsilon>0$ it is positive-recurrent, i.e. the mean first return time to the origin is finite. Therefore, for these values of $\beta$ and $\epsilon$ we can state that the Lamperti parameter $\delta$, which characterizes the distribution of the occupation time of the positive axis and the distribution of the number of the visits at the starting site (see section \ref{Sec:Distr:Occ_Visits}), is equal to $1$. Instead for $\epsilon<0$ we have $\delta=0$.

For $\beta>1$ one has that the process is null-recurrent for all $\epsilon$, namely it is recurrent and the mean return time is infinite. However, for the time being, we are not able to determine $\delta$ in this case.

Now let us consider the continuum limit of this model. The master equation is the same of \eref{Pdf:master:eq} with the new definitions \eref{Gilli:Gen:1} and \eref{Gilli:Gen:2} for the transition probabilities. After the substitutions $x=j\delta x$, $t=j\delta t$ and $P_n(j)=p(x,t)\delta x$, we get:
\begin{equation}
\fl p(x,t+\delta t)=\frac{1}{2}\left(1-\textrm{sgn}(x)\frac{\epsilon \delta x^\beta}{|x-\delta x|^\beta} \right) p(x-\delta x,t) +\frac{1}{2}\left(1+\textrm{sgn}(x)\frac{\epsilon \delta^\beta}{(x+\delta x)^\beta} \right) p(x+\delta x,t).
\end{equation}
By Taylor expanding the quantities above one has
\begin{equation}
\frac{\partial p(x,t)}{\partial t}=\frac{\delta x^2}{\delta t}\left[\frac{1}{2}\frac{\partial^2 p(x,t)}{\partial x^2}+\textrm{sgn}(x)\cdot \epsilon \delta x^{\beta-1}\frac{\partial}{\partial x}\left(\frac{p(x,t)}{|x|^\beta} \right) \right].
\end{equation}
To obtain the continuum limit we have to take $\delta x,~\delta t\to 0$, but first we need to distinguish the cases $\beta<1$ and $\beta>1$.

\begin{figure}[h!!!]
        \centering
        \includegraphics[scale=0.4]{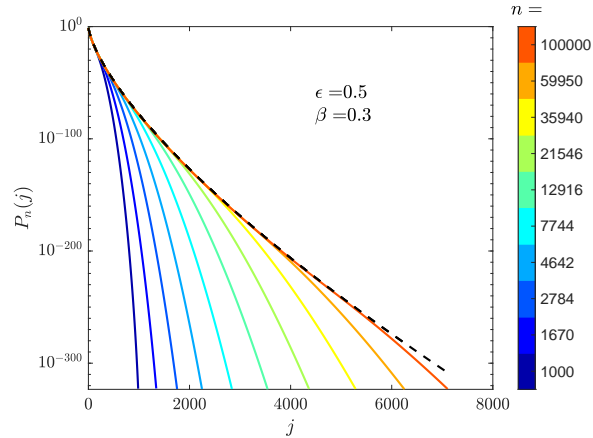}
\caption{The figure presents the probability $P_n(j)$ depending both on the position $j$ (x-axis) and the number of steps (different lines colours) for $\beta=0.3$ and $\epsilon=0.5$. The black dashed line corresponds to the stationary solution \eref{Gillis:mod:stationary} given by the continuum limit.}\label{Fig:Gillis:mod:stationary}
\end{figure}

When $\beta<1$, the term $\delta x^{\beta-1}$ diverges in the limit $\delta x\to 0$. Thus, besides keeping $\delta x^2/\delta t=D$,  we impose that the product $\epsilon \delta x^{\beta-1}$ remains constant and equal to $\varepsilon$. Consequently, to get the continuum limit in which $\delta x$, $\delta t \to 0$, we have to consider also the limit $\epsilon\to 0$.  We highlight that in the original model we did not need to consider the last limit due to the fact that for $\beta=1$ the diverging term $\delta x^{\beta-1}$ disappears. Finally, the diffusion equation is
\begin{equation}\label{Gillis:mod:ergodic}
\frac{\partial p}{\partial t}=\frac{D}{2}\frac{\partial^2 p(x,t)}{\partial x^2}+\textrm{sgn}(x)\cdot D\varepsilon\frac{\partial}{\partial x}\left(\frac{p(x,t)}{|x|^\beta} \right)
\end{equation}
and it corresponds to the Fokker-Plank equation of a diffusive particle subject to a power-law force decreasing with distance as $|x|^{-\beta}$. We observe that eq.\eref{Gillis:mod:ergodic} admits, for $\epsilon>0$, the stationary solution
\begin{equation}
p_s(x)=\frac{1}{\Gamma\left(1+\frac{1}{1-\beta} \right)} \left(\frac{2^\beta\varepsilon}{1-\beta} \right)^{\frac{1}{1-\beta}}\exp\left(-\frac{2\varepsilon |x|^{1-\beta}}{1-\beta} \right)
\label{Gillis:mod:stationary}
\end{equation}
and, consequently, the process is ergodic, as we have already stated above by Lamperti criteria.  Moreover, since the stationary solution as $x$ increases decays as a stretched exponential, all the moments $\langle |x|^q \rangle_t$  tends asymptotically to a constant, while in the original model in the ergodic regime, see \eref{spectrum:ergodic}, we obtained that only the lowest ones tend to a constant. In fig.\ref{Fig:Gillis:mod:stationary} we present the probability $P_n(j)$ of being at site $j$ after $n$ steps for $\beta=0.3$ and $\epsilon=0.5$ and show that it tends to the stationary solution \eref{Gillis:mod:stationary}. On the contrary, the stationary solution has no physical meaning  for $\epsilon<0$ and, therefore, the process is clearly non ergodic. Indeed, we have seen before that in this case it is transient.

For $\beta>1$, we have that $\delta x^{\beta-1}$ vanishes in the limit $\delta x\to 0$. Consequently,  we still have to keep constant the product $\delta x^{\beta-1} \epsilon$, but now we should consider $\epsilon\to\infty$ as $\delta\to 0$, which is obviously impossible. In fact, this requirement would imply transitions probabilities greater than $1$ or smaller than $0$. Therefore, in the continuum limit the term containing the bias vanishes and the resulting equation is simply that of a freely diffusing particle. This fact is not surprising: since for $\beta > 1$ the drift term is fast-decaying with the distance from the origin, we expect that asymptotically the evolution will be effectively described by pure diffusion, without any bias. 

We underline that the results obtained in this section are in agreement with the ones obtained in \cite{Bray}, where there are considered random walks subject to a force of the form $F(x)\sim x^{-\sigma}$, and the two systems are equivalent just setting $\beta=\sigma$.

\section{Conclusions}\label{Sec:6}
We have considered many facets of Gillis model, which represents an outstanding solvable model of a non-homogeneous random walk.

It turns out that such a model both exhibits subtle mathematical features describing multiple regimes as the single parameter it contains is varied, and as well it is a discrete realization of a diffusing particle in the presence of a logarithmic potential.

In particular, when the parameter is tuned such that a stationary probability measure exists, transport properties and record statistics exhibit peculiar, highly non-trivial features. As non-homogeneous stochastic processes represent a strongly physically motivated scenarios for which few general results are known, we both remark the outstanding presence of an exactly solvable model, and hope that further statistical properties may be analysed for stochastic systems lacking translational invariance.

\section*{Acknowledgements}
The authors gratefully thank the referees for their comments which helped improve the paper and  acknowledge partial support from PRIN Research Project No. 2017S35EHN “Regular and stochastic behavior in dynamical systems” of the Italian Ministry of Education, University and Research (MIUR).

\appendix
\section{Generating function of the probability of being at the origin }\label{App_A}
We illustrate a procedure (see \cite{Hughes_RWRE1}) more general than the one introduced in \cite{Gillis} to obtain the solution \eref{Gillis:P(z)} of the $1$-d Gillis model. More precisely, we consider the generating function $P(z|j_0)$ of the probability $P_n(j_0)$ that the walker is at the origin by starting at a generic site $j_0$
\begin{equation}
P(z|j_0)=\sum_{n=0}^\infty P_n(j_0) z^n;
\end{equation}
the generating function used in the main text is thus $P(z)=P(z|j_0=0)$. We start from the Chapman Kolmogorov equation of the propagator:
\begin{equation}
P_{n+1}(j|j_0)=\mathcal{R}(j-1) P_n(j-1|j_0)+\mathcal{L}(j+1) P_n(j+1|j_0),
\end{equation}
where $P_n(j|j_0)$ is the probability of being at site $j$ having started in $j_0$ after $n$ steps. By multiplying both sides by $z^{n+1}$ and summing over $n$ we obtain an identity involving generating functions
\begin{eqnarray}
 P(z;j|j_0)-&\delta_{j,j_0}=\frac{z}{2}\left[P(z;j+1|j_0)+P(z;j-1|j_0) \right] \\
&+\frac{\epsilon z}{2(j+1)}P(z;j+1|j_0)-\frac{\epsilon z}{2(j-1)}P(z;j-1|j_0).\nonumber
\end{eqnarray}
We reexpress such an identity in terms of the Fourier transform
\begin{equation}
\hat{P}(z;q|j_0)=\sum_{j=-\infty}^{+\infty} e^{iqj} P(z;j|j_0),
\end{equation}
obtaining 
\begin{equation}
\frac{1-z \cos q}{\sin q} \hat{P}(z;q|j_0)-\frac{e^{iqj_0}}{\sin q}=-i\epsilon z \sum_{j\neq 0} \frac{e^{iqj}}{j}P(z;j|j_0).
\end{equation}
By taking the derivative with respect to $q$ of both sides we get a differential equation for $\hat{P}$ in the form:
\begin{eqnarray}
 \partial_q \hat{P}(z;q|j_0)&+\frac{\sin q}{1-z\cos q}\left[\frac{d}{dq}\left(\frac{1-z\cos q}{\sin q}\right)- \epsilon z \right]\hat{P}(z;q|j_0) \\
&+\frac{\epsilon z \sin q}{1-z\cos q} P(z|j_0)-\frac{d}{dq}\left(\frac{e^{iqj_0}}{\sin q}\right)\frac{\sin q}{1-z\cos q}=0.\nonumber
\end{eqnarray}
The formal solution of this equation reads:
\begin{eqnarray}
\hat{P}(z;q|j_0)&=\frac{c\sin q}{(1-z\cos q)^{1-\epsilon}}+\frac{\sin q}{(1-z\cos q)^{1-\epsilon}}\int dq \frac{d}{dq}\left( \frac{e^{iqj_0}}{\sin q}\right)\frac{1}{(1-z\cos q)^\epsilon} \nonumber\\
&-P(z|j_0)\epsilon z\frac{\sin q}{(1-z\cos q)^{1-\epsilon}}\int dq \frac{1}{(1-z\cos q)^\epsilon}
\end{eqnarray}
where the constant $c$ is a function of $z$, $j_0$ and $\epsilon$.

We can now obtain $P(z|j_0)$ by using the inversion formula for the discrete Fourier transform. Integrating from $0$ to $2\pi$ we get, after integrating by parts and rearranging terms
\begin{equation}
P(z|j_0)=\frac{\int_{0}^{2\pi} e^{iqj_0}(1-z\cos q)^{-1-\epsilon}dq}{\int_{0}^{2\pi} (1-z\cos q)^{-\epsilon}dq}.
\end{equation}
The integrals can be evaluated in terms of hypergeometric functions, yielding
\begin{equation}
P(z|j_0)=\frac{z^{|j_0|}}{|j_0|!}\frac{\Gamma(1+\epsilon+|j_0|)}{2^{|j_0|}\Gamma(1+\epsilon)}\frac{_2F_1\left(\frac{1+\epsilon+|j_0|}{2},\frac{\epsilon+j_0}{2}+1;|j_0|+1;z^2\right)}{_2F_1\left(\frac{1}{2}\epsilon,\frac{1}{2}\epsilon+\frac{1}{2};1;z^2 \right)},
\label{Gillis:P(z|k0)}
\end{equation}
which is the generalization of \eref{Gillis:P(z)} derived by Gillis: such an expression is simply obtained by putting $j_0=0$ in \eref{Gillis:P(z|k0)}.

\section{Form of the generating function of the return probability}\label{App_B}
To demonstrate eq.\eref{P(z):Form} let us recall the exact form of $P(z)$:
\begin{equation}
P(z)=\frac{_2 F_1\left(\frac{1}{2}\epsilon+1,\frac{1}{2}\epsilon+\frac{1}{2};1;z^2\right)}{_2 F_1\left(\frac{1}{2}\epsilon,\frac{1}{2}\epsilon+\frac{1}{2};1;z^2\right)}.
\label{Gillis:P(z):App}
\end{equation}
We observe that, being $P(z)$ a function of $z^2$, we can consider the following generating function
\begin{equation}
\Pi(z)=P(\sqrt{z})=\sum_{n=0}^\infty \pi_n z^n=\frac{_2 F_1\left(\frac{1}{2}\epsilon+1,\frac{1}{2}\epsilon+\frac{1}{2};1;z\right)}{_2 F_1\left(\frac{1}{2}\epsilon,\frac{1}{2}\epsilon+\frac{1}{2};1;z\right)},
\label{Gillis:P(z):App}
\end{equation}
where the $n$th coefficient $\pi_n$ corresponds to $P_{2n}$. In this way we can use the transformation formulas \cite{Abr-Steg}:
\begin{eqnarray}
\fl _2 F_1\left(a,b;c;z\right)=&\frac{\Gamma(c)\Gamma(c-a-b)}{\Gamma(c-a)\Gamma(c-b)}~ _2F_1\left(a,b;a+b-c+1;1-z\right) \\
&+(1-z)^{c-a-b} \frac{\Gamma(c)\Gamma(a+b-c)}{\Gamma(a)\Gamma(b)} ~_2F_1\left(c-a,c-b;c-a-b+1;1-z\right),\nonumber
\label{2F1:1}
\end{eqnarray}
when $c-a-b$ is non-integer, while for the integer case we use
\begin{eqnarray}
\fl _2 F_1\left(a,b;a+b+m;z\right)&=\frac{\Gamma(m)\Gamma(a+b+m)}{\Gamma(a+m)\Gamma(b+m)}
\sum_{n=0}^{m-1} \frac{(a)_n(b)_n}{n!(1-m)_n}(1-z)^n \\
&-(z-1)^m \frac{\Gamma(a+b+m)}{\Gamma(a)\Gamma(b)} \sum_{n=0}^{\infty} \frac{(a+m)_n (b+m)_n}{n!(n+m)!}(1-z)^n [\log(1-z) \nonumber\\
&-\psi(n+1)-\psi(n+m+1)+\psi(a+n+m)+\psi(b+n+m)] \nonumber
\label{2F1:2}
\end{eqnarray}
and
\begin{eqnarray}
\fl _2 F_1\left(a,b;a+b+m;z\right)&= (1-z)^{-m}\frac{\Gamma(m)\Gamma(a+b-m)}{\Gamma(a)\Gamma(b)}
\sum_{n=0}^{m-1} \frac{(a-m)_n (b-m)_n}{n!(1-m)_n}(1-z)^n \nonumber\\
&-(-1)^m \frac{\Gamma(a+b-m)}{\Gamma(a-m)\Gamma(b-m)} \sum_{n=0}^{\infty} \frac{(a)_n (b)_n}{n!(n+m)!}(1-z)^n [\log(1-z)\nonumber
\\
&-\psi(n+1)-\psi(n+m+1)+\psi(a+n)+\psi(b+n)]
\label{2F1:3}
\end{eqnarray}
for $m=1,2,\cdots$, or
\begin{eqnarray}
\fl _2 F_1\left(a,b;a+b;z\right)=&\frac{\Gamma(a+b)}{\Gamma(a)\Gamma(b)} \sum_{n=0}^{\infty} \frac{(a)_n (b)_n}{(n!)^2}(1-z)^n [2 \psi (n+1) \nonumber\\
&-\psi (a+n)-\psi (b+n)-\log (1-z)],
\label{2F1:4}
\end{eqnarray}
where $\psi(z)=\frac{d}{dz}\log \Gamma(z)$ and $(z)_n$  denote respectively the digamma function and the Pochhammer's symbol \cite{Abr-Steg}.  At his point, making the substitutions \eref{2F1:1}-\eref{2F1:4}, we obtain the following form for $\Pi(z)$:

\paragraph{For $-1<\epsilon<-\frac{1}{2}$}
\begin{equation}
\Pi(z)=G\left(\frac{1}{1-z}\right),
\end{equation}
where the slowly varying function $G(x)$ is given by
\begin{equation}
\fl G(x)=a_1 \frac{ _2F_1 \left(\frac{1}{2}\epsilon+1,\frac{1}{2}\epsilon+\frac{1}{2};\frac{3}{2}+\epsilon,\frac{1}{x} \right) + a_2 x^{1/2+\epsilon}~_2F_1\left( -\frac{1}{2}\epsilon,\frac{1}{2}-\frac{1}{2}\epsilon;\frac{1}{2}-\epsilon;\frac{1}{x}\right)~}
{_2F_1 \left(\frac{1}{2}\epsilon,\frac{1}{2}\epsilon+\frac{1}{2};\frac{1}{2}+\epsilon,\frac{1}{x} \right)+a_3 x^{-1/2+\epsilon}~_2F_1\left( 1-\frac{1}{2}\epsilon,\frac{1}{2}-\frac{1}{2}\epsilon;\frac{3}{2}-\epsilon;\frac{1}{x}\right)}.
\end{equation}
The numerical coefficients $a_1$, $a_2$ and $a_3$, which depend on $\epsilon$, can be determined from \eref{2F1:1}.
\paragraph{For $\epsilon=-\frac{1}{2}$} 
\begin{equation}
\Pi(z)=G\left(\frac{1}{1-z}\right),
\end{equation}
but in this case $G(x)$ has the expression
\begin{equation}
\fl G(x)=\frac{ \sum_{n=0}^\infty\frac{(3/4)_n (1/4)_n }{(n!)^2} x^{-n}\left[2\psi(n+1)-\psi(\frac{3}{4}+n)- \psi(\frac{1}{4}+n)+\log(x)\right] }
{4+\frac{1}{4}\sum_{n=0}^\infty \frac{(3/4)_n (5/4)_n }{n! (n+1)!} x^{-n-1} \left[ \log(x)+\psi(n+1)+\psi(n+2)-\psi(\frac{3}{4}+n)-\psi(\frac{5}{4}+n) \right]}.
\end{equation}
\paragraph{For $-\frac{1}{2}<\epsilon<\frac{1}{2}$}
\begin{equation}
\Pi(z)=\frac{1}{(1-z)^{1/2+\epsilon}}G\left(\frac{1}{1-z}\right),
\end{equation}
where $G(x)$ is given by
\begin{equation}
\fl G(x)=b_1 \frac{ _2F_1\left( -\frac{1}{2}\epsilon,\frac{1}{2}-\frac{1}{2}\epsilon;\frac{1}{2}-\epsilon;\frac{1}{x}\right) + b_2 x^{-1/2-\epsilon}~_2F_1 \left(\frac{1}{2}\epsilon+1,\frac{1}{2}\epsilon+\frac{1}{2};\frac{3}{2}+\epsilon,\frac{1}{x} \right)}
{_2F_1 \left(\frac{1}{2}\epsilon,\frac{1}{2}\epsilon+\frac{1}{2};\frac{1}{2}+\epsilon,\frac{1}{x} \right)+b_3 x^{-1/2+\epsilon}~_2F_1\left( 1-\frac{1}{2}\epsilon,\frac{1}{2}-\frac{1}{2}\epsilon;\frac{3}{2}-\epsilon;\frac{1}{x}\right)}.
\end{equation}
The numerical coefficients $b_1$, $b_2$ and $b_3$, which depend on $\epsilon$, can be determined from \eref{2F1:1}.
\paragraph{For $\epsilon=-\frac{1}{2}$}
\begin{equation}
\Pi(z)=\frac{1}{1-z}G\left(\frac{1}{1-z}\right),
\end{equation}
with $G(x)$ 
\begin{equation}
\fl G(x)=\frac{4-\frac{1}{4}\sum_{n=0}^\infty \frac{(3/4)_n (5/4)_n }{n! (n+1)!} x^{-n-1} \left[ \log(x)+\psi(n+1)+\psi(n+2)-\psi(\frac{3}{4}+n)-\psi(\frac{5}{4}+n) \right]}
{ \sum_{n=0}^\infty\frac{(3/4)_n (1/4)_n }{(n!)^2} x^{-n}\left[2\psi(n+1)-\psi(\frac{3}{4}+n)- \psi(\frac{1}{4}+n)+\log(x)\right] }.
\end{equation}
\paragraph{For $\frac{1}{2}<\epsilon<1$}
\begin{equation}
\Pi(z)=\frac{1}{(1-z)}G\left(\frac{1}{1-z}\right),
\end{equation}
where $G(x)$ has the expression
\begin{equation}
\fl G(x)=c_1 \frac{ _2F_1\left( -\frac{1}{2}\epsilon,\frac{1}{2}-\frac{1}{2}\epsilon;\frac{1}{2}-\epsilon;\frac{1}{x}\right) + c_2 x^{-1/2-\epsilon}~_2F_1 \left(\frac{1}{2}\epsilon+1,\frac{1}{2}\epsilon+\frac{1}{2};\frac{3}{2}+\epsilon,\frac{1}{x} \right)}
{_2F_1\left( 1-\frac{1}{2}\epsilon,\frac{1}{2}-\frac{1}{2}\epsilon;\frac{3}{2}-\epsilon;\frac{1}{x}\right)+c_3 x^{1/2-\epsilon}~_2F_1 \left(\frac{1}{2}\epsilon,\frac{1}{2}\epsilon+\frac{1}{2};\frac{1}{2}+\epsilon,\frac{1}{x} \right)}
\end{equation}
The numerical coefficients $c_1$, $c_2$ and $c_3$, which depend on $\epsilon$, can be determined from \eref{2F1:1}.

Finally, to demonstrate eq.\eref{P(z):Form}, one obtains that if $\Pi(z)$ has the form
\begin{equation}
\Pi(z)=\frac{1}{(1-z)^\nu}G(\frac{1}{1-z}),
\end{equation}
with $G(z)$ a slowly varying function, then also $P(z)$ is of the same form
\begin{equation}
P(z)=\frac{1}{(1-z)^\nu}H(\frac{1}{1-z}),
\end{equation}
where $H(x)$ is still a slowly varying function connected to $G(x)$ by
\begin{equation}
H(x)=\frac{x^\nu}{(2x-1)^\nu}G\left(\frac{x^2}{2x-1}\right).
\end{equation}
Therefore $P(z)$ and $\Pi(z)$ share the same form with also the same parameter $\nu$.

\section{First return probability} \label{App_C}
As we have seen in the main text, due to the form \eref{F(z):form} of the generating function $F(z)$, it is not possible to use directly the Tauberian theorem to find the asymptotic behaviour of the first return probability $F_n$.  The ploy is to consider the derivative of $F(z)$, which for $\nu\neq 0$ is
\begin{equation}
F'(z)=\nu(1-z)^{\nu-1}L\left( \frac{1}{1-z}\right)-(1-z)^{\nu-2}L'\left(\frac{1}{1-z} \right).
\end{equation}
From the Lamperti theorem, see eq.\eref{Lamp:delta_1}, we can state that the following holds:
\begin{equation}
\lim_{z\to 1^-} \frac{(1-z)^{-1} L'\left(\frac{1}{1-z}\right)}{L\left(\frac{1}{1-z}\right)}=0.
\end{equation}
Consequently, for $\nu\neq 0$, the leading order term is
\begin{equation}
F'(z)\sim \nu(1-z)^{\nu-1}L_0\left(\frac{1}{1-z}\right),
\end{equation} 
where $L_0(x)$ is the dominating term of $L(x)$. For $\nu=0$ the derivative is simply given by
\begin{equation}
F'(z)=-\frac{1}{(1-z)^2}L'\left( \frac{1}{1-z}\right),
\end{equation}
but in this case generally $L'(x)$ is not slowly-varying.

As an example let us consider $\epsilon=-\frac{1}{2}$, in this case $H_0(x)=\frac{1}{4}\log(x)$ and then 
\begin{equation}
-L'(x)\sim \frac{4}{x \log^2(x)},
\end{equation}
so we have
\begin{equation}
F'(z)\sim -\frac{4}{(1-z)\log^2(1-z)}.
\end{equation}
Now, considering the definition of the generating function $F(z)$ its derivative is written as
\begin{equation}
F'(z)=\sum_{n=0}^\infty n F_n z^{n-1},
\end{equation}
so, via the Tauberian theorem for power series \cite{Feller}, we obtain the behaviour of the mean recurrence time 
\begin{equation}
\tau_n=\sum_{k=1}^n k F_k\sim \frac{n}{\log^2(n)}.
\end{equation}
Moreover, assuming that the sequence $\lbrace 2nF_{2n}\rbrace$ is (ultimately) monotonic, we can also state that 
\begin{equation}
F_{2n}\sim\frac{4}{n\log^2(n)}.
\end{equation}
For the other cases we obtain
\begin{equation}
F_{2n}\sim \cases{
n^{-(1/2-\epsilon)} ~~~&for $-1<\epsilon<-\frac{1}{2}$\\
\frac{1}{n\log^2(n)} &for $\epsilon=-\frac{1}{2}$\\
n^{-(3/2+\epsilon)}~~~&for $-\frac{1}{2}<\epsilon<1$\\
}
\label{1D:Fn:exp}
\end{equation}
and, see also \cite{Hughes-1986},
\begin{equation}
\tau_n\sim \cases{
n^{3/2+\epsilon} ~~~&for $-1<\epsilon<-\frac{1}{2}$\\
\frac{n}{\log^2(n)} &for $\epsilon=-\frac{1}{2}$\\
n^{1/2+\epsilon}~~~&for $-\frac{1}{2}<\epsilon<\frac{1}{2}$\\
\log(n), &for $\epsilon=\frac{1}{2}$\\
\frac{2\epsilon}{2\epsilon-1}&for $\frac{1}{2}<\epsilon<1$.
}
\label{tau:app}
\end{equation}
\section{Distribution of the maximum of an excursion}\label{App_D}
Here we consider the method of the Lyapunov functions, see \cite{Mens-Pop-Wade}, to obtain the distribution $\mathcal{M}_m(x)$ of the excursion maximum $m$. This method consists in finding a function $\mathcal{F}(x)$ of a stochastic process $X_s$ such that its image $Y_s=\mathcal{F}(X_s)$ has specific characteristics, through which it is possible to determine the properties of the original process.

In our case we want that $Y_s$ has the property that the first moment of the increment $\bar{
\Delta}_s=Y_{s+1}-Y_s$ is equal to zero for all $s$. To this aim let us consider
\begin{equation}
Y_s = X_s^\gamma,
\label{Lyap:Function}
\end{equation}
with $\gamma>0$, and evaluate the increment of the new process:	
\begin{eqnarray}
\bar{\Delta}_s&= (X_s+\Delta_s)^\gamma-X_s^\gamma=X_s^\gamma\left[ \left(1+\frac{\Delta_s}{X_s}\right)^\gamma-1 \right] \nonumber\\ &\approx \gamma \Delta_s X_s^{\gamma-1}+\frac{\gamma(\gamma-1)}{2}  \Delta_s^2 X_s^{\gamma-2}
\end{eqnarray} 
where we have considered the Taylor expansion up to second order in $\Delta_s= X_{s+1} - X_s$. Now, to find $\gamma$ such that $Y_s$ has no drift, let us consider the expectation of $\bar{\Delta}_s$
\begin{eqnarray}
\mathbb{E}(\bar{\Delta}_s | X_s=x) &\approx \gamma  \mathbb{E}(\Delta_s | X_s=x) x^{\gamma-1}+ \frac{\gamma(\gamma-1)}{2}  \mathbb{E}(\Delta_s^2 | X_s=x) x^{\gamma-2}\nonumber\\
&=\gamma \left( -\epsilon + \frac{\gamma-1}{2}	\right)   x^{\gamma-2},
\label{Delta:bar_s}
\end{eqnarray}
in the last equality we used \eref{mu:1} and \eref{mu:2} for $\mathbb{E}(\Delta_s | X_s=x)$ and  $\mathbb{E}(\Delta_s^2 | X_s=x)$. Finally, imposing that \eref{Delta:bar_s} is equal to zero, we obtain
\begin{equation}
\gamma=1+2\epsilon.
\label{gamma:epsilon}
\end{equation}
Consequently, the random walk defined in \eref{Lyap:Function} with this value for $\gamma$  corresponds to a symmetric random walk. At this point, for the stochastic process $Y_s$, one has \cite{Redner}:
\begin{equation}
\mathbb{P}\left( Y_s~\mathrm{hits}~y~\mathrm{before~reaching}~0~|~Y_0=y_0 \right)=\frac{y_0}{y}.
\end{equation}
Therefore, making use of eq.\eref{Lyap:Function}, we finally get
\begin{equation}
\mathbb{P}\left( X_s~\mathrm{hits}~x~\mathrm{before~returning~to~}0\right)\sim\frac{1}{x^{\gamma}}.
\end{equation}
We observe, as done in the main text, that if a particle, which begins its motion at the origin, hits the position $x$ before returning to the starting point, then the maximal position reached by the particle must be greater than $x$. Thus the former relation implies:
\begin{equation}
\mathbb{P}\left( m\geq x \right)\sim\frac{1}{x^{\gamma}},
\end{equation}
from which we obtain the form of the distribution $\mathcal{M}_m(x)$:
\begin{equation}
\mathcal{M}_m(x)=\mathbb{P}(m<x)=1-C x^{-\gamma}.
\end{equation}

\end{document}